\documentclass[11pt]{article}
\usepackage{pdfsync}
\usepackage{graphicx} 
\usepackage{subfig}
\usepackage{color}
\usepackage{amsfonts}
\usepackage{amssymb}
\usepackage{epsfig}
\usepackage{dcolumn}
\usepackage{booktabs}
\usepackage{multirow}
\usepackage{makecell}
\usepackage{boldline}
\usepackage[colorlinks]{hyperref}
\hypersetup{
    colorlinks=true,
    linkcolor=blue,
    filecolor=magenta,  
    urlcolor=[rgb]{0.00,0.00,0.50},    
    citecolor=red,
    }

\usepackage{url}

\advance \textheight by \topskip
\usepackage{amsmath}
\usepackage[english]{babel}
\makeatletter
 \@addtoreset{equation}{section}
\makeatother

\usepackage{amsmath,amssymb}
\makeatletter \@addtoreset{equation}{section}
\makeatother
\def\be{\begin{equation}}

\def\be{\begin{equation}}
\def\ee{\end{equation}}

\def\cW{{\mathcal{W}}}

\def\Z{\mathbb Z}
\def\R{\mathbb R}

\def\C{\mathbb C}

\usepackage{amsmath,amssymb}
\def\bea{\begin{eqnarray}}
\def\eea{\end{eqnarray}}
\def\barray{\begin{array}}
\def\earray{\end{array}}

\def\sn{\mathrm{sn}}
\def\cn{\mathrm{cn}}
\def\dn{\mathrm{dn}}

\def\nd{\mathrm{nd}}
\def\cd{\mathrm{cd}}
\def\sd{\mathrm{sd}}

\def\cA{\mathcal{A}}

\def\cW{\mathcal{W}}

\begin{document}
%%%%%%%%%%%%%%%%%%%%%%%%%%%%%

\title{
{\bf Position-dependent mass,  finite-gap systems, and supersymmetry
}}

\author{{\bf Rafael Bravo  and Mikhail S. Plyushchay}  \\
[8pt]
{\small \textit{
Departamento de F\'{\i}sica,
Universidad de Santiago de Chile, Casilla 307, Santiago 2,
Chile  }}\\
[4pt]
 \sl{\small{E-mails: 
\textcolor{blue}{rafael.bravog@usach.cl}, 
\textcolor{blue}{mikhail.plyushchay@usach.cl}
}}
}
\date{}
\maketitle

\begin{abstract}
The ordering problem in quantum systems with 
position-dependent mass (PDM)  is treated
 by  inclusion of the classically fictitious
 similarity  transformation into the kinetic term.
This provides a generation of 
supersymmetry  with the first order supercharges
from the kinetic term alone, while 
inclusion of the potential term allows  
also
 to generate 
nonlinear supersymmetry with higher order supercharges.
A broad class of finite-gap systems with PDM
is obtained by different reduction procedures,
and general  results on supersymmetry generation
are applied to them.
We show that elliptic finite-gap systems 
of Lam\'e and Darboux-Treibich-Verdier types
can be obtained by reduction to  
Seiffert's spherical spiral and Bernoulli lemniscate in the presence of
Calogero-like or harmonic oscillator potentials, or 
by angular momentum reduction of a free motion 
 on some 
$AdS_2$-related surfaces 
in the presence of Aharonov-Bohm flux. 
The limiting cases include
the Higgs and Mathews-Lakshmanan 
oscillator models  as well as a reflectionless 
model with PDM exploited recently in the discussion of cosmological 
inflationary scenarios.

\end{abstract}

\vskip.5cm\noindent

\section{Introduction}
Quantum mechanical systems with position-dependent mass (PDM) appear 
in physics in various contexts. 
When a particle interacts with an external environment, 
its mass is replaced by an effective mass that in general 
depends on the position. As a result, quantum systems with PDM emerge
naturally in solid state physics  where heterostructures are characterized  by 
electrons' effective masses \cite{solid1,Roos,Harrison,Morrow,Bast}.
In another way,  they can be  generated
via a dimensional reduction of 
field-theoretical nonlinear sigma models 
and  in a related framework of gravitation 
\cite{Susskind,Speight,Gukov,QuesTkach,Harko}. 
A certain class of such systems is used particularly  in 
cosmological inflationary models \cite{Linde}.
Quantum mechanical  systems with PDM  were 
employed recently in the context of integrable models \cite{GangDas}.
They turn out to be interesting from the point of view of 
supersymmetric quantum mechanics 
\cite{susy0,susy1,susy2,susy3,susy4}, coherent states
 \cite{coherent1,coherent2,coherent3},
and PT-symmetry \cite{PT0,PT1,PT2}.
Besides  the one-dimensional quantum systems with PDM, 
their multi-dimensional  generalizations are considered in the literature 
\cite{D0,D1,D2,D3,D3+,D4},
particularly, in the context of  superintegrable systems \cite{BalFra,Nikitin}.
See also refs. 
\cite{Levy1,Roy,Alhaid,Koc,Gang1,Gang2,Schulze,Gang3,CCNN,Midya,Levai,Mazh1,Killi} 
where some other aspects of quantum systems with PDM were
studied.

In treating quantum systems with PDM,
there appears the ordering problem  in the kinetic term.
One can take a classical analog of such a system, remove 
the position dependence in the kinetic term via appropriate
point (canonical) transformation, and then  quantize 
the obtained system  with translation-invariant
kinetic term, the transformed potential term and,
possibly, changed domain of the transformed coordinate variable.
When working with such  systems,
however, usually consideration starts directly at the quantum
level by choosing some fixed ordering prescription in the kinetic term,
or by considering  some family of orderings. This picture 
with the  two possibilities to start from  the classical or quantum levels 
is somewhat reminiscent of the Dirac's dilemma in quantization
of constrained systems: ``first reduce and then quantize" or  
``first quantize and then reduce" \cite{Dir1,Dir2,Dir3}.

In this paper we analyze the problem of the quantum ordering 
in the kinetic term with PDM  in one dimension in a 
new
 way
which turns out having a 
certain analogy 
with  the 
treatment of the quantum problem of a particle in 
a curved space \cite{deWitt,deAlf}.
For this we  introduce a kind of a similarity transformation 
in a kinetic term.
\emph{Classically}
 such a transformation is 
\emph{artificial and fictitious}, 
but
its direct quantum  analog is nontrivial and 
\emph{reflects
effectively  the  quantum ordering ambiguity }
in the kinetic term with a position dependent mass. 
This will allow us to incorporate in a simple way 
supersymmetry into the framework of the 
one-dimensional quantum mechanical
 models with PDM. 
The general results we obtain are applied 
then to a broad class of finite-gap quantum elliptic systems 
of the Lam\'e and Darboux-Treibich-Verdier types, and to  their 
limiting cases  with a single real or hidden imaginary period.
The systems with PDM we consider belong to a class of 
nonlinear dynamical systems of Li\'enard type \cite{Lienard}, 
for which on concrete examples we observe the  peculiarities
associated with the presence of poles in mass function.
We also show how the corresponding 
finite-gap systems can be obtained by different 
reduction procedures from either a free particle 
motion  on some  surfaces of revolution, namely, 
on $AdS_2$, sphere $S^2$ or on $AdS_3$,
or by appropriate reduction of the particles moving in  Euclidean $\R^2$,
Minkowski $\R^{1,1}$, or spherical $S^2$ spaces in the presence of 
the Calogero- or harmonic oscillator-type potentials.
In this way,  finite-gap elliptic systems are obtained
via reduction to  Seiffert's spherical  spiral and  Bernoulli's 
lemniscate (in a special case of the modular parameter value),
or by angular momentum reduction 
of a free motion on certain  $AdS_2$-related surfaces
in the presence of Aharonov-Bohm flux.
We show  that supersymmetric pairs of finite-gap systems
related by the first order intertwining operators 
are generated naturally from the kinetic term  
with the PDM only,  
\emph{without a necessity 
of  introducing apart of a potential term}.
The inclusion of the potential term 
allows us to extend the construction 
for the case of supersymmetry based on the  
\emph{ higher order 
differential  intertwining  operators}. 
 
The paper is organized as follows. 
 In Section \ref{FicSUSY} we start from the observation 
 how supersymmetric pairs of the systems 
 can be  generated 
 by quantization of a kinetic term with 
 PDM
  into which
 a fictitious  classical similarity transform 
 is introduced.
 In Section \ref{Kinm(x)} we  show  
 that the  inclusion of the classically fictitious function
into the kinetic term with PDM allows us  
to transfer the ordering ambiguities under transition to the quantum case
into the similarity transform 
function while keeping  fixed the position of the 
PDM function. 
In such a way we 
 \emph{ cover universally all the
distinct ordering prescriptions}
 in the kinetic term 
with PDM  considered
in  literature, and show that 
the construction of Section  \ref{Kinm(x)} corresponds 
to a particular choice of the ordering prescription for
position-dependent mass function $m(x)$ of arbitrary form.
Section \ref{SecFinPDM} is devoted to the discussion
of some models of finite-gap systems with PDM.
Namely, we consider there some  finite-gap families of 
hyperbolic reflectionless, trigonometric 
and elliptic systems of the Lam\'e and Darboux-Treibich-Verdier
types.  We discuss the relation between the indicated 
families of the systems, consider 
peculiarities of their phase space trajectories 
associated with a presence 
of the poles in mass function,  discuss shortly quantum
properties of the systems, and consider different 
reduction procedures by which 
the systems can be generated.
In Section \ref{SecSUSY} we apply 
general results of Sections  \ref{FicSUSY} 
and  \ref{Kinm(x)} to generate supersymmetric 
extentions of the families of finite-gap systems 
from Section \ref{SecFinPDM}. 
The last Section \ref{LastSec} is devoted to 
concluding remarks and discussion
of some interesting problems for future research.

%%%%%%%%%%%%%%
%%%%%%%%%%%%%%%%%%%%%%%%%%%%%

\section{Supersymmetry from a
fictitious similarity transform}\label{FicSUSY}

Consider a free non-relativistic particle of mass $M=1/2$ in one dimension. 
Its classical kinetic term $h_1=p^2$ 
can be 
written  
in an equivalent  form 
\be\label{p2eta}
	h_{\varsigma}=\varsigma(x)\,	p \frac{1}{\varsigma^2(x)}p\,\varsigma(x)=
	\left(-i\varsigma(x)\,	p \frac{1}{\varsigma^(x)}\right)
	\left(i\frac{1}{\varsigma^(x)} p \varsigma(x)\right)
\ee
with arbitrary real-valued function   $\varsigma(x)$ which we restrict by 
the condition  $\varsigma(x)>0$.
Since $h_{\varsigma}=h_1$, \emph{classically}  
 dependence  of $h_{\varsigma}$ on $\varsigma(x)$   is fictitious. 
This observation  can be  generalized further by taking, e.~g.,
\bea
h_{\varsigma_1,\varsigma_2,\varsigma_3;\alpha}&=&
\alpha \varsigma_1(x)\,	p \frac{1}{\varsigma_1^2(x)}p\,\varsigma_1(x)
\nonumber
\\
&& +\,\frac{1}{2}(1-\alpha)\left(\varsigma_2(x)\,	p \frac{1}{\varsigma_2(x)\varsigma_3(x)}p\,\varsigma_3(x)
\,+\, 
 \varsigma_3(x)\,	p \frac{1}{\varsigma_3(x)\varsigma_2(x)}p\,\varsigma_2(x)
\right)\,,\label{gen1}
\eea
where $\alpha$ 
  is a real constant and $\varsigma_a(x)>0$, $a=1,2,3$, are arbitrary functions.

\vskip0.1cm

Quantum analog of 
(\ref{p2eta}) depends     on the choice of
$\varsigma(x)$ as well as 
on the ordering prescription for non-commuting factors. 
Let us  take~\footnote{We use the  units with $\hbar=1$; the Planck
constant will be restored where necessary.} 
\be\label{Bsigma}
	A_\varsigma=\frac{1}{\varsigma(x)}\frac{d}{dx}
	\varsigma(x)=\frac{d}{dx}+W(x)\,,\qquad 
	W(x)=\frac{d}{dx}\ln \varsigma(x)=
\frac{\varsigma'(x)}{\varsigma(x)}\,,
\ee
as  a quantum analog of the classical term
$i\frac{1}{\varsigma(x)}p\,\varsigma(x)$
which appears on the right in (\ref{p2eta}).
We also have 
\be\label{Bsigma+}
	A_{1/\varsigma}=-A^\dagger_\varsigma=\varsigma(x) 
	\frac{d}{dx}\frac{1}{\varsigma(x)}=\frac{d}{dx}-W(x)\,
\ee
as a quantum analog (with a minus sign)  of another factor 
$-i\varsigma(x) p\frac{1}{\varsigma(x)}$ in (\ref{p2eta}).
Then the direct quantum analog of   (\ref{p2eta})
can be presented in a factorized form,
\bea\label{Heta}
	H_\varsigma&=&-\varsigma(x)\,	\frac{d}{dx}\frac{1}{\varsigma^2(x)}
	\frac{d}{dx}\,\varsigma(x)\\
\label{Heta+}
	&=& A^\dagger_\varsigma A_\varsigma=-\frac{d^2}{dx^2}+
	W^2-W'\,.
\eea

The  factorizing first order 
differential operators  (\ref{Bsigma}), (\ref{Bsigma+}) and Hamiltonian  (\ref{Heta})
are invariant under scaling transformations 
$\varsigma\rightarrow e^C \varsigma$,
$C\in\R$.
On the other hand, 
the inversion $\varsigma\rightarrow 1/\varsigma$ induces 
the interchange of  $A_{\varsigma}$ and $A_{\varsigma}^\dagger$ : 
 $A_{1/\varsigma}=-A^\dagger_\varsigma$,
$A^\dagger_{1/\varsigma}=-A_\varsigma$.
This generates a  permutation  of non-commuting 
operators in (\ref{Heta+}), 
\be\label{H1/eta}
	H_{1/\varsigma}=A^\dagger_{1/\varsigma} A_{1/\varsigma}= 
	A_\varsigma A^\dagger_\varsigma=-\frac{d^2}{dx^2}+
	W^2+W'\,.
\ee
Operators $A_\varsigma$ and $A^\dagger_\varsigma$ intertwine the 
Hamiltonians (\ref{Heta}) and (\ref{H1/eta}), $A_\varsigma H_\varsigma= 
H_{1/\varsigma} A_\varsigma$,
$A^\dagger_\varsigma H_{1/\varsigma}=  H_\varsigma A^\dagger_\varsigma$, and so,
generate the Darboux transformation \cite{Matveev} 
between 
the systems given  by the quantum 
Hamiltonians (\ref{Heta}) and (\ref{H1/eta}).
In a usual way, one can compose a  $2\times 2$ matrix Hamiltonian operator
$\mathcal{H}_\varsigma=\text{diag}\,(H_{\varsigma},H_{1/\varsigma})$
and obtain   the  $N=2$ supersymmetric system with
supercharges $\mathcal{Q}_{\varsigma +}=
A_\varsigma^\dagger \sigma_+$ and 
$\mathcal{Q}_{\varsigma -}=\mathcal{Q}_{\varsigma +}^\dagger=
A_\varsigma  \sigma_-$, $\sigma_\pm=\frac{1}{2}(\sigma_1
\pm i \sigma_2)$,
constructed from the 
intertwining operators $A_\varsigma $ and $A_\varsigma^\dagger$.
The operators $\mathcal{H}_\varsigma$ and  $\mathcal{Q}_{\varsigma \pm}$ 
generate 
the $N=2$ supersymmetry, 
$[\mathcal{H}_\varsigma, \mathcal{Q}_{\varsigma \pm}]=0$, 
$\mathcal{Q}_{\varsigma \pm}^2=0$, 
$[\mathcal{Q}_{\varsigma +},\mathcal{Q}_{\varsigma -}]_{{}_+}=
\mathcal{H}_\varsigma$, in which  the diagonal Pauli matrix 
$\sigma_3$ 
plays  a role  of the $\Z_2$ grading operator 
\cite{Witten,SUSYQM}.

\vskip0.1cm

One can also choose a more general ordering by taking
$H_{\varsigma,\alpha}=\frac{1}{2}(1+\alpha) H_\varsigma + 
\frac{1}{2}(1-\alpha) H_{1/\varsigma}$.
Then 
\be\label{Hsigalp} 
	H_{\varsigma,\alpha}=-\frac{d^2}{dx^2}+W^2-  \alpha W'\,,\qquad
	H_{1/\varsigma,\alpha}=H_{\varsigma,-\alpha}\,.
\ee
This corresponds to  a direct quantum analog of the classical 
expression in  (\ref{gen1}) with $\varsigma_1=\varsigma$, $\varsigma_2=\varsigma_3=1/\varsigma$
and with $\alpha$ changed for $\frac{1}{2}(1+\alpha)$.
We shall see that   a pair ($H_{\varsigma,\alpha}$, $H_{\varsigma,-\alpha}$) also 
can be  associated with supersymmetry. 

The peculiarity of the quantum systems (\ref{Heta+}), 
(\ref{H1/eta}) and (\ref{Hsigalp} ) 
is that if we restore the Planck constant $\hbar$ in them, we obtain
that their corresponding induced potential terms are proportional  
to $\hbar^2$.  We shall discuss this point 
later.

\vskip0.1cm

Till the moment, the introduction of $\varsigma(x)$ starting from the 
classical kinetic term (\ref{p2eta}) with $M=1/2$ seems 
to be rather artificial.
Below  we pass over to  the case of the position-dependent mass,
where $\varsigma(x)$ transforms into a natural 
element of the construction.

\section{Kinetic term with a
PDM and supersymmetry}
 \label{Kinm(x)}

Consider now a one-dimensional  system
described by Lagrangian with a position-dependent mass $M(x)\equiv \frac{1}{2}m(x)>0$,
\be\label{tildeL}
	L(x)=\frac{1}{4}m(x){\dot{x}^2}-
	u(x)\,.
\ee
In the changed notation for the mass, the case $m=1$ corresponds to 
$M=1/2$, and in what follows we shall refer to $m(x)$ as a mass.
The Euler-Lagrange equation of motion for (\ref{tildeL})
can be presented in the form 
\be\label{ddotx}
	{\ddot{x}}=-2\frac{u'(x)}{m(x)} 
	-\frac{1}{2}\frac{m'(x)}{m(x)}\dot{x}^2\,.
\ee
Equation  (\ref{ddotx}) corresponds to a class of 
nonlinear dynamical systems of Li\'enard type, namely, 
of the  quadratic type with the dynamics given by the equation
of the  form  $\ddot{x}+f(x)\dot{x}^2+g(x)=0$ \cite{Lienard}.

Let us denote $f(x)=\frac{1}{\sqrt{m(x)}}$, and 
make a point transformation  $x\rightarrow \chi$ with 
$d\chi=\frac{dx}{f(x)}$.
 Then
\be\label{xi(x)}
	\chi=\chi(x)=\int^x \frac{d\eta}{f(\eta)}=\int^x \sqrt{m(\eta)}\,d\eta\,.
\ee
The inverse to (\ref{xi(x)}) transformation is  
\be\label{xchi}
\qquad 
		x=x(\chi)=\int^\chi \varphi(\eta)d\eta\,,
\ee
where  
\be\label{varphifx}
	 \varphi(\chi)=f(x(\chi))\,.
\ee
Using $d\chi=\frac{dx}{f(x)}$, one can rewrite Lagrangian (\ref{tildeL})  in the form 
\be\label{Lchi}
	L(\chi,x)=\frac{1}{4}\dot{\chi}^2-U(\chi)-\frac{1}{2}
	\kappa(x-x(\chi))^2\,,
\ee
where $x(\chi)$ is given by  (\ref{xchi}),  $\kappa\neq 0$ is a constant,
and $U(\chi)=u(x(\chi))$.
The Euler-Lagrange equations for (\ref{Lchi}) are
(i) $x=x(\chi)$, 
and (ii) $\ddot{\chi}=-2U'(\chi)$.
Equation (i) yields 
$\dot{\chi}=\dot{x}/f(x)$, and  
then from (ii) we obtain 
\be\label{ddotx+}
	\ddot{x}=-2u'(x)f^2(x) +\frac{f'(x)}{f(x)}\dot{x}^2\,,
\ee
that is equivalent to (\ref{ddotx}). 
Changing   
 $\frac{1}{2}\kappa(x-\mu(\chi))$
for a  
Lagrange multiplier $\lambda$, one can obtain 
an equivalent  to (\ref{Lchi}) form of Lagrangian, 
$
	L(\chi,x,\lambda)=\frac{1}{4}\dot{\chi}^2-U(\chi)
	-\lambda
	(x-x(\chi))
	$.
	
 \vskip0.1cm

A canonical transformation 
$(x,p)\rightarrow (\chi,P)$ with $P=f(x) p$ corresponds to the 
point transformation 
(\ref{xi(x)}). It transforms the Hamiltonian $h_{m(x)}=\frac{1}{m(x)}p^2+
u(x)$ of the system (\ref{tildeL}) with position-dependent mass
into the Hamiltonian  $h_1=P^2+U(\chi)$
with  $m=1$.

\vskip0.1cm

The Hamiltonian   
kinetic term $h_{m(x)}=\frac{1}{m(x)}p^2$
with position-dependent mass 
can be presented 
 in an 
equivalent  symmetric 
 form similarly to (\ref{p2eta}),
\be\label{Hm(x)cl}
	h_{f,\varsigma}=f(x)\varsigma(x)p
	\frac{1}{\varsigma^2(x)}p\varsigma(x)f(x)=
	\left(-if(x)\varsigma(x)p
	\frac{1}{\varsigma(x)}\right)\left(i
	\frac{1}{\varsigma(x)}p\varsigma(x)f(x)\right)
	\,.
\ee
This  can be considered as a classical kinetic term 
with a position-dependent mass $m(x)=1/(f(x))^2$ 
 and   function
$\varsigma(x)$ of a  fictitious similarity transform.
As a quantum analog of (\ref{Hm(x)cl})  we take
\be\label{Hetaf}
	H_{f,\varsigma}=-f\varsigma \frac{d}{dx} \frac{1}{\varsigma^2} \frac{d}{dx}
	\varsigma f
	=A^\dagger_{f,\varsigma}A_{f,\varsigma}\,,
\ee
where 
\be\label{Bsigmaf}
	A_{f,\varsigma}=\frac{1}{\varsigma}
	\frac{d}{dx}\varsigma f\,,\qquad
	A^\dagger_{f,\varsigma}=-f\varsigma
	\frac{d}{dx}\frac{1}{\varsigma}=-A_{f,1/{\varsigma f}}
	\,.
\ee

The ordering in Eqs. (\ref{Bsigmaf}) and (\ref{Hetaf}) is chosen
in such a way  that 
for $f(x)=1$ they  reduce to
(\ref{Bsigma}) and (\ref{Bsigma+}), (\ref{Heta}).
Then,  for 
\vskip0.1cm

({\bf i}) $\varsigma=1$
\vskip0.1cm

\noindent 
 but with nontrivial $f(x)$, we have 
\be\label{Hfsi0}
	H_{(0)}=-f  \frac{d^2}{dx^2}f, 
\ee
that reproduces the kinetic term for a system 
with position-dependent mass $m(x)=1/(f(x))^2 $. 
Such ordering prescription 
was considered, e.g.,   in \cite{Roos,Roy,Killi}. 
A more general choice of
\vskip0.1cm

({\bf ii})  $\varsigma=m^{\nu+\frac{1}{2}}$

\vskip0.1cm
\noindent 
 in (\ref{Hetaf}) 
yields the kinetic term $-m^\nu \frac{d}{dx} m^{-2\nu-1} \frac{d}{dx} m^\nu$ 
of a form which was considered in \cite{Roos,Morrow}. 
For $\nu=-\frac{1}{2}$, we reproduce (\ref{Hfsi0}).
The case $\nu=0$ corresponds to \cite{Levy1,Roy,Killi}
\be\label{Hfsi1}
	H_{(1)}=- \frac{d}{dx}f^2\frac{d}{dx}\,.
\ee
The choice $\nu=-\frac{1}{4} $ yields
\be\label{Hfsi1/2}
	H_{(1/2)}=- f^{1/2}\frac{d}{dx}f\frac{d}{dx}f^{1/2}\,.
\ee
The  origin of the notation for  lower index in $H$ in (\ref{Hfsi0}), (\ref{Hfsi1}) and (\ref{Hfsi1/2}) 
will be clarified below.
Kinetic terms of the form \cite{Levy1,Roy,Killi}

\vskip0.1cm
({\bf iii}) $-\frac{1}{2}(f^2\frac{d^2}{dx^2}+\frac{d^2}{dx^2}f^2)$, 
\vskip0.1cm

\noindent 
and  \cite{Killi}
\vskip0.1cm

({\bf iv}) $-\frac{1}{2}(\frac{d}{dx}f\frac{d}{dx}f+
f\frac{d}{dx}f\frac{d}{dx})$, 
\vskip0.1cm

\noindent 
which represent particular cases of a  generalized form for 
the quantum kinetic term 
\be\label{fag}
H_{\alpha,\beta,\gamma}=-\frac{1}{2} 
\left(f^\alpha \frac{d}{dx} f^\beta \frac{d}{dx} f^\gamma +
f^\gamma \frac{d}{dx} f^\beta \frac{d}{dx} f^\alpha\right)
\ee
with  
$\alpha+\beta+\gamma=-2$ \cite{Roos,Levy1},
 are also included in 
(\ref{Hetaf}) for particular choices of $\varsigma(x)$, see Appendix~A.
\vskip0.1cm
Thus the inclusion of the classically fictitious function $\varsigma$
into the kinetic term allows us  
to transfer the ordering ambiguities under transition to the quantum case
into now a true similarity transform  
function $\varsigma$ while keeping  fixed the position of the function
$f(x)=1/\sqrt{m(x)}$ in quantum kinetic term  (\ref{Hetaf}). 
In such a way we  cover 
all the 
distinct ordering prescriptions in the kinetic term 
with PDM considered
in the literature.
Moreover,  this
also  gives us a possibility to treat  distinct ordering prescriptions in a 
unified way.
\vskip0.1cm

Consider now a similarity transformation generated by 
the function 
$f^{1/2}=m^{-1/4}(x)$. 
We have 
\be
	f^{1/2}A_{f,\varsigma} f^{-1/2}=
	\frac{1}{f^{1/2}\varsigma}\left(f\frac{d}{dx}\right)
	(f^{1/2}\varsigma)\,,\qquad
	f^{1/2}A^\dagger_{f,\varsigma} f^{-1/2}=-(f^{1/2}\varsigma)
	\left(f\frac{d}{dx}\right)\frac{1}{f^{1/2}\varsigma}\,.	
\ee
Denoting 
$
\Sigma(\chi)=\varsigma(x)\vert_{x=x(\chi)}\,,
$
one gets
\be 
	f^{1/2}A_{f,\varsigma} f^{-1/2}\vert_{x=x(\chi)}=
	\cA_{\Phi}\,,\qquad
	f^{1/2}A^\dagger_{f,\varsigma} f^{-1/2}\vert_{x=x(\chi)}=
	\cA^\dagger_{\Phi}\,,
\ee	
where 
\be
	\cA_{\Phi}=\frac{1}{\Phi(\chi)}\frac{d}{d\chi}\Phi(\chi)=\frac{d}{d\chi}
	+\cW(\chi)\,,\qquad
	\cA^\dagger_{\Phi}=-{\Phi(\chi)}\frac{d}{d\chi}\frac{1}{\Phi(\chi)}=
	-\cA_{1/{\Phi}}\,,
\ee
and 
\be\label{superpot}
	\cW(\chi)=\frac{d}{d\chi}\ln \Phi(\chi)=
	\frac{\Phi'(\chi)}{\Phi(\chi)}\,.
\ee
The kernel $\Phi(\chi)$ of the first order operator  
$\cA_{\Phi}^\dagger$    is given by 
\be\label{Phisxigen}
	\Phi(\chi)=\varphi^{1/2}(\chi)\Sigma(\chi)\,,
\ee
while the   kernel of  $\cA_{\Phi}$ is $1/\Phi(\chi)$.
For the similarity-transformed Hamiltonian (\ref{Hetaf}) we have 
the chain of equalities
\be\label{HPhi}
	f^{1/2}H_{f,\varsigma}f^{-1/2}\vert_{x=x(\chi)}=
	\cA^\dagger_{\Phi} \cA_{\Phi}=
	-\Phi(\chi)\frac{d}{d\chi}\frac{1}{\Phi^2(\chi)}\frac{d}{d\chi}
	\Phi(\chi)=-\frac{d^2}{d\chi^2}+\cW^2-\cW'\equiv H_\Phi\,.
\ee
After similarity transformation and the change of variable
the quantum Hamiltonian (\ref{Hetaf}) takes exactly  the form
of the quantum kinetic term  (\ref{Heta}) but with 
the  Darboux generating function 
 $\varsigma(x)$ changed for $\Phi(\chi)$.
\vskip0.1cm

Consider a special family of 
the  functions 
\be\label{sigflam}
\varsigma=f^{-\lambda}
\ee
 given 
in terms of position-dependent mass
 that corresponds to  the ordering 
 ({\bf ii})
considered above with $\lambda=2\nu+1$. 	
In this case $\Phi(\chi)$ from  (\ref{Phisxigen}) reduces to
\be\label{Phix1/2}
	\Phi(\chi)\rightarrow (\varphi(\chi))^{\frac{1}{2}-
	\lambda}\equiv\Phi_{(\lambda)}(\chi)\,.
\ee
Let us denote the corresponding operators  also 
by the lower index $(\lambda)$. Then  for a particular value $\lambda=1/2$ we have 
the first order operators
\be
	A_{(1/2)}=f^{1/2}\frac{d}{dx}f^{1/2}=-A_{(1/2)}^\dagger\,
\ee
which factorize  the quantum kinetic term (\ref{Hfsi1/2}), $H_{(1/2)}=
A_{(1/2)}^\dagger A_{(1/2)}$.
Since
$\Phi_{(1/2)}(\chi)=1$,  the similarity-transformed operators reduce to
\be\label{H0free} 
	f^{1/2}A_{(1/2)}f^{-1/2}=f\frac{d}{dx}=\frac{d}{d\chi}\,,\qquad
	f^{1/2}H_{(1/2)}f^{-1/2}=-\frac{d^2}{d\chi^2}\,.
\ee
This corresponds to a free particle with $\chi$ taking values 
in the domain which is defined by  the domain of the initial position 
variable $x$ as well as by the  form of  the position-dependent
mass function $m(x)$. 
Therefore, for any position dependent mass $m(x)$, 
there is a special choice (\ref{Hfsi1/2}) of ordering
in the kinetic term, which after similarity transformation and 
change of variable reduces the kinetic term to the form of  
the quantum kinetic term 
( \ref{H0free}) with $m=1$ and $\chi$ taking values in the corresponding domain.

For $\lambda=1$ we obtain the quantum kinetic operator (\ref{Hfsi1}), which 
factorizes as 
\be
	H_{(1)}=A_{(1)}^\dagger A_{(1)}\,,\qquad
	A_{(1)}=f\frac{d}{dx}\,,\qquad
	A_{(1)}^\dagger=-\frac{d}{dx} f\,.
\ee
On the other hand,
the choice  $\lambda=0$ yields the quantum operator  (\ref{Hfsi0})
for which we have 
\be
	H_{(0)}=A_{(0)}^\dagger A_{(0)} \,,\qquad
	A_{(0)}=\frac{d}{dx}f\,,\qquad
	A_{(0)}^\dagger=-f\frac{d}{dx} \,.
\ee	
Since  $\Phi_{(0)}(\chi)=(\varphi(\chi))^{1/2}=
1/{\Phi_{(1)}(\chi)}$, then
$A_{(1)}^\dagger=-A_{(0)}$, and
$H_{(1)}$ and $H_{(0)}$ form a pair of super-partners
intertwined by $A_{(0)}$ and $A_{(0)}^\dagger=
-A_{(1)}$ :
$A_{(0)}H_{(0)}=H_{(1)}A_{(0)}$, 
$H_{(0)}A_{(0)}^\dagger=A_{(0)}^\dagger H_{(1)}$.

In a similar way, a pair of the similarity-transformed 
Hamiltonians $H_{(\lambda_1)}$ and 
$H_{(\lambda_2)}$
with $\lambda_1+\lambda_2=1$ after the change of variable
$x\rightarrow \chi$  takes a standard form of 
a pair of super-partner Schr\"odinger 
 Hamiltonians.  The explicit form of a one-parameter family of 
 supersymmetric pairs of kinetic Hamiltonian operators 
 with position-dependent mass is 
 \be\label{Hlam-lam}
 	H_{(\lambda)}=f^{1-\lambda}\frac{d}{dx}f^{2\lambda}\frac{d}{dx}f^{1-\lambda}\,,
	\qquad
	H_{(1-\lambda)}=f^{\lambda}\frac{d}{dx}f^{2-2\lambda}\frac{d}{dx}f^{\lambda}\,.
 \ee

In generic case, if two quantum systems are given by 
the pairs of functions ($f_1(x)$,  $\varsigma_1(x)$)
and ($f_2(x)$,  $\varsigma_2(x)$) such that
$f_1\varsigma_1^2=Cf_2\varsigma_2^2$,
where $C> 0$ is an arbitrary constant, 
and the domain of $\chi$ in both cases is the same,
then $\Phi_1(\chi)=C\Phi_2(\chi)$, and corresponding 
quantum systems are equivalent.

If the pairs of the functions ($f_1(x)$,  $\varsigma_1(x)$)
and ($f_2(x)$,  $\varsigma_2(x)$) are such that
$f_1\varsigma_1^2=C/(f_2\varsigma_2^2)$
and, again, the domain of $\chi$ in both cases is the same,
then  $\Phi_1(\chi)=C/\Phi_2(\chi)$, and corresponding 
Hamiltonians yield a  pair of super-partner systems.

With
 $x\in (x_1,x_2)$,
 the following equality is valid
\be\label{scalardef}
 \langle\Psi_1\vert\Psi_2\rangle \equiv \int_{x_1}^{x_2}\Psi^*_1(x)\Psi_2(x)dx=
\int_{\chi_1}^{\chi_2}\tilde{\Psi}_1^*(\chi)\tilde{\Psi}_2(\chi)d\chi\equiv 
\langle\tilde{\Psi}_1\vert\tilde{\Psi}_2\rangle\,
\ee
for a scalar product of two 
wave functions, 
where 
\be\label{psitilde}
	\tilde{\Psi}(\chi)=f^{1/2}(x)\Psi(x)\vert_{x=x(\chi)}\,.
\ee
For any differential operator ${O}(x)$,
define 
\be\label{Oxchi}
	{\mathcal{O}}(\chi)=
	f^{1/2}(x){O}(x)f^{-1/2}(x)\vert_{x=x(\chi)}\,.
\ee
Then we get
\be\label{Oxchi++}
	 \langle\Psi_1\vert {O}\vert\Psi_2\rangle =
	 \langle\tilde{\Psi}_1\vert {\mathcal{O}}\vert
	 \tilde{\Psi}_2\rangle\,.
\ee
\vskip0.1cm
The similarity transformation (\ref{Oxchi}), (\ref{psitilde})   
accompanied by the change 
of variable (\ref{xi(x)}) maps the quantum 
system (\ref{Hetaf}) given only by the kinetic term with a 
 position-dependent mass
into the system (\ref{HPhi})
with position-independent mass and a nontrivial potential term.
The correspondence between the two systems 
is established by the relation
(\ref{Oxchi++}). Again, as in the case $f=1$ considered in the preceding section,
the peculiarity of the system (\ref{HPhi})
is that if we restore the Planck constant $\hbar$, we obtain
\be\label{pothbar}
	H_\Phi=-\hbar^2\frac{d^2}{d\chi^2}+\hbar^2(\cW^2-\cW')\,.
\ee
In this case the generated potential term is proportional to $\hbar^2$
and has  a purely quantum nature.

\section{Finite-gap systems with position-dependent mass
}\label{SecFinPDM}

To apply the general results on position-dependent mass
we discussed till the moment, 
below we consider some  families of finite-gap  and reflectionless
systems.  The latter case can be considered as a corresponding 
limit of finite-gap systems with valence bands degenerating 
(after possible merging and shrinking \cite{CJNP,CJPFin}) into
the bound states \cite{PAN}. All such systems are intimately related 
to nonlinear integrable systems and are characterized by 
the presence in them of a nontrivial Lax-Novikov integral of motion.
All they are described  by potentials to be quadratic in Planck 
constant~$\hbar$.

\subsection{
General picture}
\label{subGeneral}

For a quantum system with position dependent mass,
  the ordering  (\ref{Hfsi1/2}) is special.  
In this case after a  similarity transformation and change 
of variable (\ref{xi(x)}), 
any quantum system  $H(x)=-(\sqrt{f}\frac{d}{dx}\sqrt{f})^2+u(x)$ 
 with PDM $m(x)=1/f^2(x)$ and 
potential  $u(x)$  
transforms into the quantum 
system with Hamiltonian of the standard form with
position-independent mass $m=1$, 
$H(x)\rightarrow H(\chi)=-\frac{d^2}{d\chi^2}+U(\chi)$
with $U(\chi)=u(x(\chi))$.

We consider now some  examples  of the 
systems with 
PDM
belonging to an important 
class of finite-gap systems closely related with  integrable systems, 
which find diverse interesting applications in physics.
They are presented  in Tables 
 \ref{Table1}, \ref{Table2}, \ref{Table3} below,
 which include the families of hyperbolic, {\bf H}, 
 trigonometric, {\bf T}, and elliptic,
 {\bf L} and {\bf D},  systems of such a nature. 
 Namely,  the quantum systems with position-independent mass $m=1$
 presented by the cases {\bf \, H${}_\text{1}$}, {\bf \, T${}_\text{1}$},
 {\bf \, L${}_\text{1}$} and {\bf \, D${}_\text{1}$}
 with  potentials of the form $u_{{}_{C_n}}(\chi)=C_n u_1(\chi)$
 are finite-gap for $C_n=n(n+1)\hbar^2$, $n=1,2,\ldots$. 
 Each such a quantum system
 $H(\chi)=-\hbar^2\left(\frac{d^2}{d\chi^2}+n(n+1)u_1(\chi)\right)$
 possesses 
 a nontrivial Lax-Novikov integral which is a  differential operator  of order $2n+1$.
 Let us  stress that for  finite-gap systems potential term includes the 
 multiplicative quantum factor  $\hbar^2$, cf. (\ref{pothbar}).
 In the cases  {\bf \, H${}_\text{1}$} and  {\bf \, L${}_\text{1}$}
 Lax-Novikov operators  are the true integrals of motion being analogs of the 
 free particle momentum operator of the zero-gap case $n=0$. 
 The systems  from  the family {\bf \, L${}_\text{1}$} 
 are the quantum $n$-gap Lam\'e systems with periodic (elliptic)
 potential~\footnote{
 The dependence of Jacobi's elliptic functions on modular parameter
 $k$, $0<k<1$, is not shown explicitly here; $k'=\sqrt{1-k^2}$ denotes
 the complementary modular parameter, $\text{\bf K}=\text{\bf K}(k)$ 
 is the complete elliptic integral of the first kind, 
 and $\text{\bf K}'=\text{\bf K}(k')$
 \cite{WW,NIST}.
 We indicate the dependence on modular parameter 
 explicitly where it will be necessary.}  
 $u_{{}_{C_n}}(\chi)=-C_n\, \text{dn}^2\,\chi$. 
 In the infinite-period limit corresponding to $k\rightarrow 1$,
 $n$ valence bands shrink and transform into $n$ bound states 
 of a reflectionless  
 system belonging to the class of the hyperbolic P\"oschl-Teller 
 systems with potential $u_{{}_{C_n}}(\chi)=-C_n\, \frac{1}{\cosh^2\,\chi}$.
The Lax-Novikov integral in the $n$-gap Lam\'e quantum system detects 
all the edge states of the continuous bands by annihilating them, and 
distinguishes the left- and right-moving 
Bloch states inside the valence and conduction bands
by the sign of  their eigenvalues \cite{PAN}.
Analogous role is played  by the Lax-Novikov integrals
in reflectionless systems, where they detect 
the bound states and the edge state of the conduction band,
and separate the 
left- and right-moving analogs of the plane waves 
in the continuous part of the spectrum. 
The systems represented by the case {\bf \, D${}_\text{1}$} 
with potentials $u_{{}_{C_n}}(\chi)=C_n\,\text{dc}^2\,\chi$
can be  obtained from the family 
{\bf \, L${}_\text{1}$} by a complex displacement 
$\chi \rightarrow 
\chi+\text{\bf K}+i\text{\bf K}'$,
which corresponds to the complex half-period of the Lam\'e potential,
accompanied by an additive shift,  
$\text{dc}^2\,\chi=-\text{dn}^2\,(\chi+\text{\bf K}+i\text{\bf K}')+1$.
In another way, the {\bf \, D${}_\text{1}$}  family can be obtained 
from the {\bf \, L${}_\text{1}$} family by transformations
$\chi \rightarrow i\chi$, $k\leftrightarrow k'$ with subsequent 
multiplication of the Lagrangian by $(-1)$,
$L\rightarrow -L$. Analogously,
transformations $\chi\rightarrow i\chi$, $L\rightarrow -L$
produce the trigonometric family {\bf \, T${}_\text{1}$}
from the hyperbolic one {\bf \, H${}_\text{1}$}
and vice versa.
The series {\bf \, D${}_\text{1}$}   
belongs to a more broad family  of Darboux-Treibich-Verdier 
finite-gap systems with singular 
(at $\chi=\pm \text{\bf K}$) potentials \cite{Brezhnev,Veselov}. 
In the limit $k\rightarrow 0$ the {\bf \, D${}_\text{1}$} family transforms 
into the family  {\bf \, T${}_\text{1}$}
given by $u_{{}_{C_n}}(\chi)=C_n\, \frac{1}{\cos^2\,\chi}$
with $-\pi/2<\chi<\pi/2$. 
 The systems from the family  {\bf \, T${}_\text{1}$}
 are almost isospectral to  a free particle  confined 
 inside  the infinite potential well,
 and can be obtained from the latter
 by applying to it the appropriate 
 Darboux-Crum transformation of order $n$,
 like the systems of the family  {\bf \, H${}_\text{1}$}
 can be obtained by Darboux-Crum  transformations 
 from the free particle on a real line.
The systems from the family {\bf \, D${}_\text{1}$} 
can be considered as a  periodization 
in the `hidden imaginary direction'
of the systems {\bf \, T${}_\text{1}$}
like Lam\'e systems can be treated as periodicized 
in the real variable $x$
reflectionless P\"oschl-Teller systems
having a hidden imaginary period.
Unlike the cases of 
 {\bf \, H${}_\text{1}$} and {\bf \, L${}_\text{1}$}
systems, 
the Lax-Novikov operators in the families
{\bf T${}_\text{1}$} and 
 {\bf \, D${}_\text{1}$} are the formal
integrals of motion.
 Though they commute with corresponding Hamiltonian 
 operators, acting on 
 the bound states
 they produce non-physical states 
 which violate boundary 
 conditions~\footnote{Cf. this with finite-gap 
 Calogero model \cite{COP}.}. 
 
 The corresponding systems possess a series of  interesting 
 properties, which we discuss shortly below for each of the
 three families.  
 The finite-gap systems  
 with position-dependent mass are presented here
 by different special choices for  the 
 functions  $m(x)$, which are interrelated 
 in the hyperbolic, trigonometric and elliptic cases
 by the above mentioned transformations $x\rightarrow ix$,
 $L\rightarrow -L$, 
 by limit procedures $k\rightarrow 1,0$, and by periodizations. 
 The  corresponding potentials in the systems for the chosen 
 position-dependent mass functions have a form 
and nature to be very different from those 
they take after the transformation 
 $x\rightarrow \chi$.
 For instance,
the potential corresponding to the reflectionless
hyperbolic P\"oschl-Teller system, see
Table \ref{Table1}
below, 
takes there
the Calogero-like form $u_1(x)=-1/(x^2+1)$, or the 
harmonic oscillator like form $u_1(x)=x^2-1$, or the 
form of the Calogero potential transformed by 
`Zhukowsky map', $u_1(x)=-4/(x+\frac{1}{x})^2$, or the 
Mathiew (pendulum-like)  form $u_1(x)=-\cos^2 x$, or 
the elliptic generalization of the latter, $u_1(x)=-\cn^2 x$.  
We also obtain reflectionless systems 
with  potential function $u_1(x)=4(e^{-2x}-e^{-x})$ 
of the form of Morse potential.
The  classical  phase portraits for such systems have peculiarities related 
 with the presence of the real poles in the position-dependent mass.
 The  finite-gap systems  we consider  can be obtained from 
 a  particle with position-independent mass in
 Euclidean,  Minkowski, or spherical space
 in the presence of 
 Calogero-like or harmonic oscillator potential, 
 by reducing its  motion to different curves 
(which, in dependence on the case,
 can be a circle, hyperbola,  Seiffert's spherical spiral,   or 
 Bernoulli lemniscate).
 The kinetic terms of the systems from the families 
 {\bf T} and {\bf H}
 can be produced by  a reduction 
 to  geodesics on Riemann  sphere  and hyperbolic Lobachevsky
 plane as well.
 The finite-gap systems 
 can also be obtained by angular momentum reduction of  a free
 particle motion on some surfaces of revolution
 (in the presence of Aharonov-Bohm flux).

\subsection{Reflectionless systems}

Lagrangians for the systems presented in Table \ref{Table1},
$L=\frac{1}{4}m(x)\dot{x}^2-C_n u_1(x)$, 
can be obtained  by starting from  
a particle with position-independent mass 
in two-dimensional Minkowski space  which is subjected to the action 
of attractive Calogero potential, $L=\frac{1}{4}(\dot{X}^2-\dot{Y}^2) +
C_n\frac{1}{Y^2}$,
and then restricting the motion  
to the  hyperbolic curve  $X^2-Y^2=-1$.  
Six different parametrizations of  the hyperbola's branch $Y=+\sqrt{X^2+1}$
given by the functions $X(x)$ and $Y(x)$  shown in the Table  
result in six models for reflectionless systems 
presented there.
The mass function in such an interpretation 
can  be  presented initially 
in two alternative  forms
 $m(x)=X'^2/(1+X^2)=Y'^2/(Y^2-1)$, where $X'=dX/dx$.

%%%%%%%%%%%%%%%%%%%%%%%%%%%%%%%%%%%%

 \begin{table}[h]
\caption{\underline{{\bf H} -family.}\\
Corresponding  reflectionless systems 
are given by
potentials  $u_{{}_{C_n}}(x)=C_n u_1(x)$ with $C_n=n(n+1)\hbar^2$, $n=1,2,\ldots$.
 Here $\chi\in (-\infty,\infty)$; 
 $\text{gd}\,x=\text{arctan}\,(\sinh x)=
 2\,\text{arctan}\,(e^\chi)-\frac{\pi}{2}$ 
 is the gudermannian function, 
 %\textcolor{blue}{http://dlmf.nist.gov/4.23#viii}
  $\text{sn}^{-1}\,x=
 \text{arcsn}\,(x,k)$
 is the inverse to Jacobi's $\text{sn}$-function 
\cite{NIST}. 
% \textcolor{blue}{http://dlmf.nist.gov/22.15.
 }
\label{Table1}
\begin{center}
%\begin{tabular}{|c||c|c|c|c|c|c|c|c|} \hline 
\begin{tabular}{|c||c|c|c|c|cV{2.5}c|c|c|c|} \hline 
\multirow{2}{2em}{Case}
 &\multirow{2}{2em}{$m(x)$} & \multirow{2}{2em}{$\varphi(\chi)$} & 
 \multirow{2}{4em}{$x=x(\chi)$} & \multirow{2}{3em}{$(x_1,x_2)$} 
 & \multirow{2}{2em}{$u_1(x)$} & \multirow{2}{2em}{$X(x)$} & \multirow{2}{2em}{$Y(x)$} \\
 \, & \, & \, & \, & \, & \, & \, &  \\
%\toprule 
\hline
\hline
{\small  {\bf \, H${}_\text{1}$} }\, & {\small  $1$} & {\small  $1$}  & {\small  $\chi$} & 
{\small (-$\infty,\infty$)} & {\small $-\frac{1}{\cosh^2 x}$} & {\small $\sinh x$}& {\small $\cosh x$}
 \\[3pt]\hline
{\small  {\bf \, H${}_\text{a}$} }\, & 
{\small $\frac{1}{1+ x^2}$} & {\small $\cosh  \chi$}  &  
{\small $\sinh\chi$} &  {\small (-$\infty,\infty$)} & 
{\small $-\frac{1}{1+x^2}$} & {\small $x$} &{\small $\sqrt{1+x^2}$} 
 \\[3pt]\hline
{\small  {\bf \, H${}_\text{b}$} }\, & 
{\small $\frac{1}{(1-x^2)^2}$} & 
{\small $\frac{1}{\cosh^2\chi}$}  & 
{\small  $\tanh\chi$} & {\small  $(-1,1)$} & {\small $x^2-1$} & {\small $\frac{x}{\sqrt{1-x^2}}$} \ & {\small$\frac{1}{\sqrt{1-x^2}}$}
 \\[3pt]\hline
{\small  {\bf \, H${}_\text{c}$} }\, & 
{\small $\frac{1}{x^2}$} & {\small $e^{\chi}$}  &  
{\small $e^\chi$} & {\small  $(0,\infty)$} & 
{\small $- \frac{4x^2}{(1+x^2)^2}$} & 
{\small $\frac{1}{2}(x-x^{-1})$}& 
{\small $\frac{1}{2}(x+x^{-1})$} 
\\[3pt]\hline
{\small  {\bf \, H${}_\text{d}$} }\, & 
{\small $\frac{1}{\cos^2x}$} & 
{\small $\frac{1}{\cosh \chi}$}   &  
{\small $\text{gd}\,\chi$} & {\small  $(-\frac{\pi}{2},\frac{\pi}{2})$}     & {\small  $-\cos^2 x$} &
{\small $\tan x$} &  {\small $\frac{1}{\cos x}$}  
\\[3pt]\hline
{\small  {\bf \, H${}_\text{e}$} }\, &
{\small $\text{dc}^2\, x$} & {\small $\frac{1}{\sqrt{1+k'^2\sinh^2\chi}}$}  & 
{\small $\text{sn}^{-1}(\tanh \chi)$} & {\small $(-\text{\bf K},\text{\bf K})$} &
{\small $- \text{cn}^2\, x$} & {\small $\frac{\text{sn}\, x}{\text{cn}\,x}$}&   {\small $\frac{1}{\text{cn}\,x}$} 
\\[5pt] \hline

\end{tabular}

\end{center}
\end{table}

%%%%%%%%%%%%%%%%%%%%%%%%%%%%%%%%%%%%%

Kinetic terms for these {\bf H}-models 
can also be obtained from the kinetic term 
for a particle on Lobachevsky (hyperbolic)
plane by reduction  to appropriate  geodesics.
For this we can take the Poincar\'e upper half-plane model 
for Lobachevsky plane given by the metric 
$ds^2=\frac{1}{4}\frac{dX^2+dY^2}{Y^2}$, $Y>0$.
Restriction of $v^2\equiv (ds/dt)^2$
to the  geodesic $X=const$ with subsequent  change of notation $Y\rightarrow x$
yields  the position-dependent mass for the case 
{\bf  H${}_\text{c}$}.
Restriction of $v^2$ to the geodesic 
in the form of semicircle $X^2+Y^2=1$, $Y>0$,  parametrized as 
in the case {\bf  T${}_\text{1}$}, i.e., 
$X=\sin x$, $Y=\cos x$, $-\frac{\pi}{2}<x<\frac{\pi}{2}$   
(see Table \ref{Table2}
below),
yields the kinetic term corresponding to the case 
{\bf  H${}_\text{d}$}.
Restriction to the same geodesic parametrized as in the 
cases {\bf  T${}_\text{a}$}, {\bf  T${}_\text{b}$}, {\bf  T${}_\text{d}$}
and {\bf  T${}_\text{e}$} gives,
respectively, the  kinetic terms for the cases 
{\bf  H${}_\text{b}$}, {\bf  H${}_\text{a}$}, {\bf  H${}_\text{1}$} and 
{\bf  H${}_\text{e}$}.
 
Kinetic terms for hyperbolic models can be obtained 
by restriction of $v^2$ to geodesics 
in Poincar\'e disc model for Lobachevsky plane as well.
Taking   the metric 
 $ds^2=\frac{1}{4}\frac{dX^2+dY^2}{(1-X^2-Y^2)^2}$, $X^2+Y^2<1$,
and reducing it, for instance, to  a geodesic $Y=0$, $-1<X<1$,  
we generate the kinetic term for {\bf  H${}_\text{b}$}  case, etc.
 
 The case {\bf H${}_\text{e}$} in the limit $k\rightarrow 1$ transforms into
 the {\bf  H${}_\text{1}$} case, 
while in the limit $k\rightarrow 0$ it reduces to the case {\bf  H${}_\text{d}$}. 
This means that the 
 {\bf H${}_\text{e}$}  can be considered as the family 
interpolating continuously between 
the 
position-independent mass,
{\bf  H${}_\text{1}$}, and PDM, {\bf  H${}_\text{d}$}, 
cases.

The case {\bf H${}_\text{a}$} corresponds to
the Mathews-Lakshmanan `oscillator model'
\cite{Laksh}, see also  \cite{Carin}.
The equivalent form of the potential $u_1(x)$ here  is
$u_1(x)=\frac{x^2}{1+x^2}-1$, and 
up to inessential additive constant, the 
Lagrangian can be presented  in the form 
\be\label{L(1+x2)}
L=\frac{1}{1+x^2}\left(\frac{1}{4}\dot{x}^2-x^2\right)\,.
\ee
This can be considered as a zero-dimensional analogue  of 
Lagrangian density $\mathcal{L}=\frac{1}{2}\frac{1}{1+\phi^2}
(\partial_\mu \phi \partial^\mu \phi
-\phi^2)$ which appears in some nonlinear quantum field theories
\cite{Salam,Nish}.

 In the same context,
Lagrangian for the case {\bf H${}_\text{b}$}
can be treated as a zero-dimensional analogue  of 
the field Lagrangian density 
\be\label{Linde}
\mathcal{L}=\frac{1}{2}\frac{1}{(1-\phi^2)^2}
\partial_\mu \phi \partial^\mu \phi
-\frac{1}{2}\gamma \phi^2\,,
\ee
which 
was 
exploited by Linde et al 
in the discussion of cosmological inflationary scenarios \cite{Linde}. 
Below we shall return to this 
case in more detail in the context of supersymmetry.

Function $x(\chi)=\tanh \chi$ 
from   the case 
{\bf H${}_\text{b}$} describes a stationary kink solution in the 
$\varphi^4$ $(1+1)$-dimensional field model \cite{Jackiw}, and 
also appears as a solution in the Gross-Neveu model \cite{GN}.
The  function $x(\chi)=\text{gd}\,\chi$ from the
{\bf  H${}_\text{d}$} case corresponds to the kink solution 
in the sine-Gordon field theory in (1+1) dimensions
\cite{Jackiw}.

If in the case {\bf \, H${}_\text{c}$} we change 
 $m(x)=\frac{1}{x^2}$ for  $m(x)=\alpha^2\frac{1}{x^2}$ with 
 $\alpha>0$,  we obtain $f(x)=\frac{1}{\alpha}x$,  and $\chi(x)=\int^x \frac{d\eta}{f(\eta)}=
 \alpha\ln x$, $x=e^{\chi/\alpha}$, $-\infty<\chi<\infty$, $\varphi(\chi)=\frac{1}{\alpha}e^{\chi/\alpha}
 =\frac{1}{\alpha}(\cosh \frac{\chi}{\alpha}+\sinh \frac{\chi}{\alpha})$.
Then since  $\cosh \chi=\frac{1}{2}(x^{\alpha}+x^{-\alpha})$,
the potential 
\be\label{u1alpha}
u_1(x)=-\frac{4}{(x^\alpha+x^{-\alpha})^2}=
-\frac{4x^{2\alpha}}{(1+x^{2\alpha})^2}
\ee
corresponds to $U_1(\chi)=-1/\cosh^2\chi$. 
Particularly, for the choice $\alpha=1/2$,  this gives
the potential  of a simpler form $u_1(x)=-\frac{4x}{(1+x)^2}$
in comparison with the case $\alpha=1$.
This difference, however, does not produce something 
new.

As we noted, the hyperbolic family {\bf H${}_\text{1}$}
(and all other families  with position-dependent mass 
which  reduce to {\bf H${}_\text{1}$}
after similarity transformation  and the change of variable)
can be obtained by appropriate Darboux-Crum transformations from the
free particle on the real line.
In the case of the system 
with potential $U_{C_n}(\chi)$, the spectrum 
contains $n$ bound states of discrete energies $E_l=-\hbar^2 (n-l)^2$,
$l=0,\ldots, n-1$, with the  
ground state $\Phi(\chi)=(\cosh\chi)^{-n}$,
and continuous (scattering) part  with $E\geq 0$.
The Lax-Novikov integral in the case {\bf H}${}_{\bf 1}$ 
is \cite{CJPAdS}
\be\label{LaxNov}
\mathcal{P}=\mathcal{D}_{-n}\mathcal{D}_{-n+1}\ldots
\mathcal{D}_{0}\ldots \mathcal{D}_{n-1}\mathcal{D}_{n}\,,
\ee
where $\mathcal{D}_{l}=\frac{d}{d\chi}+l\tanh \chi$.
The change of variable function $\varphi(\chi)$ is related to
the ground state $\Phi(\chi)$ in a simple (exponential) way
in the cases {\bf H}${}_\text{\bf a}$, {\bf H}${}_\text{\bf b}$
and {\bf H}${}_\text{\bf d}$. By this reason, for these 
cases  it is natural to use  the ordering based on relations
(\ref{sigflam}), (\ref{Phix1/2}),
for which the Hamiltonian operator can be presented in the form
$H=-f^{1-\lambda}\frac{d}{dx}f^{2\lambda}\frac{d}{dx}f^{1-\lambda}$,
where $f(x)=(m(x))^{-\frac{1}{2}}$. 
As a result, the parameter $\lambda$ is fixed 
from relation (\ref{Phix1/2}) and takes the values
$\lambda=\frac{1}{2}-n$, 
$\lambda=\frac{1}{2}+\frac{1}{2}n$ and 
$\lambda=\frac{1}{2}+n$ in the cases
{\bf H}${}_\text{\bf a}$, {\bf H}${}_\text{\bf b}$
and {\bf H}${}_\text{\bf d}$, respectively.
The Lax-Novikov integral (\ref{LaxNov}), 
which is differential operator of order $2n+1$, 
for these cases can be presented then in the form:
\be\label{LNa}
	\text{\bf H}_{\text {\bf a}}: \quad
		\mathcal{P}=f^{n+\frac{1}{2}}\frac{d^{2n+1}}{dx^{2n+1}}f^{n+\frac{1}{2}}\,,\qquad
		f(x)=(1+x^2)^{\frac{1}{2}}\,,
\ee
\be\label{LNb}
	\text{\bf H}_{\text {\bf b}}: \,\,
		\mathcal{P}=f^{\frac{1}{2}(1-n)}\frac{d}{dx}f^{3/2}\frac{d}{dx}f^{3/2}\ldots 
f^{3/2}\frac{d}{dx}f^{3/2} \ldots 
f^{3/2}\frac{d}{dx}f^{3/2}\frac{d}{dx}f^{\frac{1}{2}(1-n)}\,,\quad
		f(x)=1-x^2\,,
\ee
\be\label{LNd}
	\text{\bf H}_{\text {\bf d}}: \quad
		\mathcal{P}=f^{\frac{1}{2}-n}\frac{d}{dx}f^{2}\frac{d}{dx}f^{2}\ldots 
f^2\frac{d}{dx}f^2\ldots 
f^2\frac{d}{dx}f^{2}\frac{d}{dx}f^{\frac{1}{2}-n}\,,\quad
		f(x)=\cos x\,.
\ee
\vskip0.1cm

\begin{figure}[htbp]
\begin{center}
\includegraphics[scale=0.42]{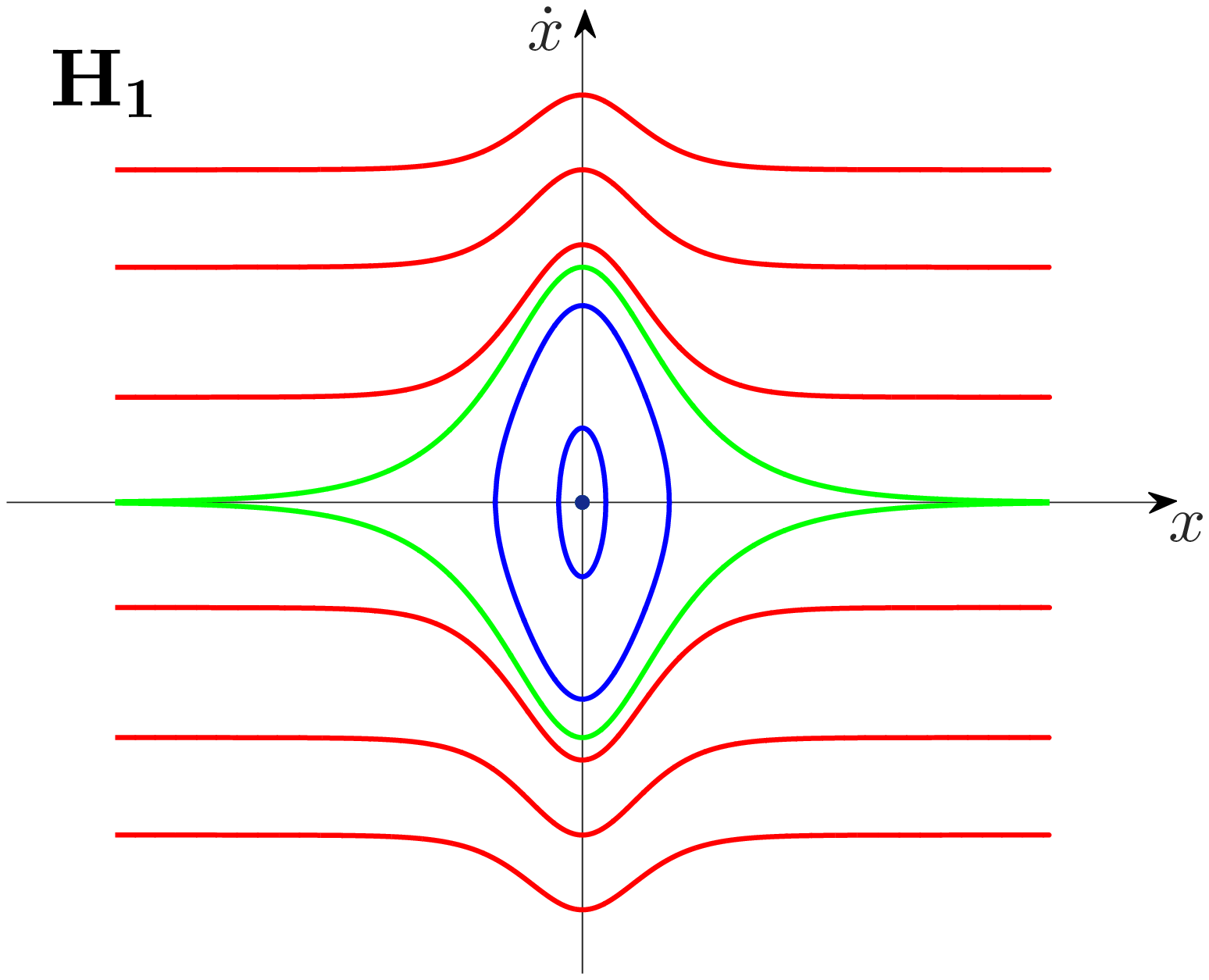} 
\includegraphics[scale=0.42]{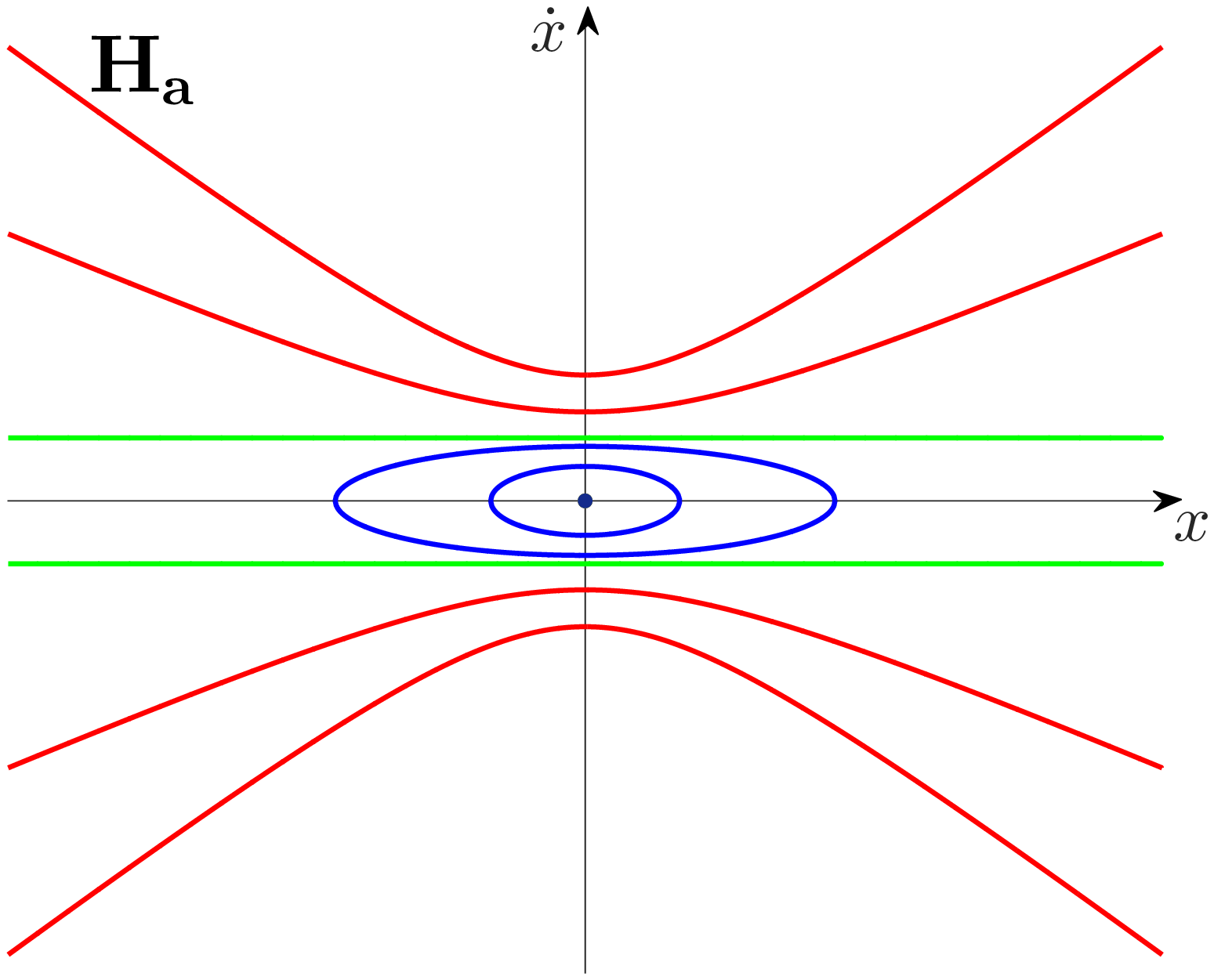}
\caption{
Phase portraits of
{\bf  H}${}_\text{\bf 1}$ and {\bf H}${}_\text{\bf a}$ systems. }
\label{HaH1}
\end{center}
\end{figure}

\begin{figure}[htbp]
\begin{center}
\includegraphics[scale=0.42]{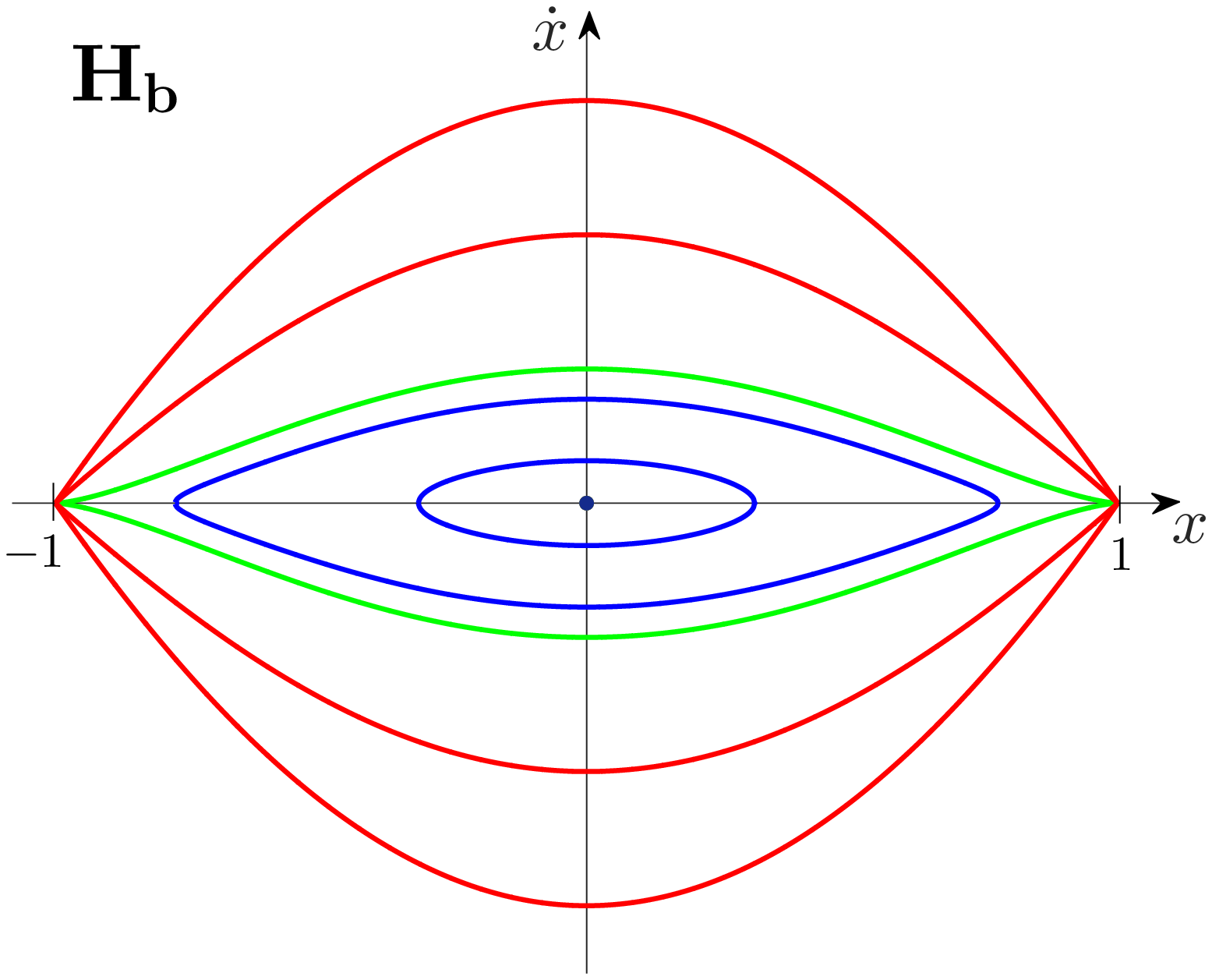}\includegraphics[scale=0.42]{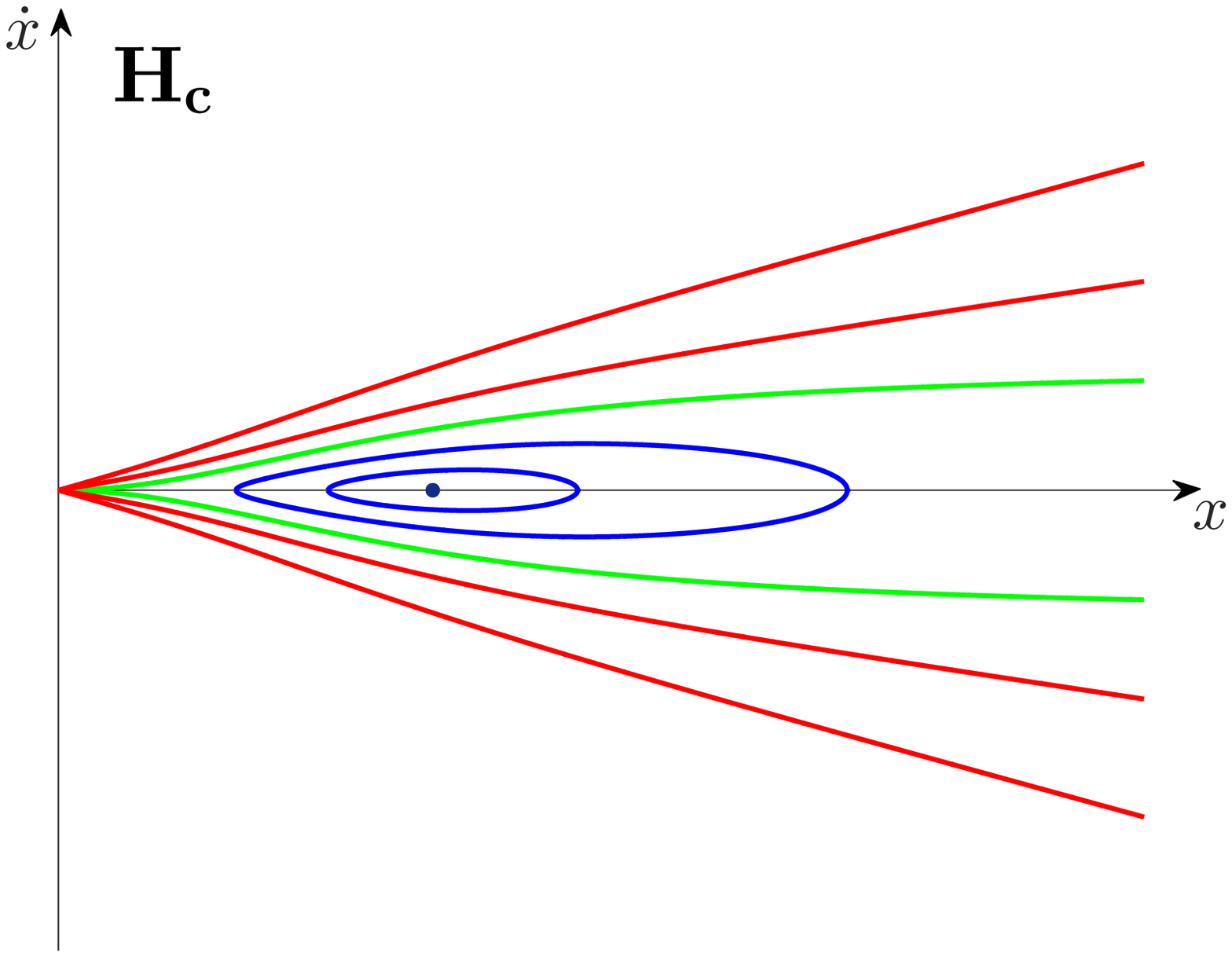}
\caption{  Phase portraits of
{\bf  H}${}_\text{\bf b}$ and {\bf H}${}_\text{\bf c}$ systems.
}\label{HbHc}
\end{center}
\end{figure}

\begin{figure}[htbp]
\begin{center}
\includegraphics[scale=0.5]{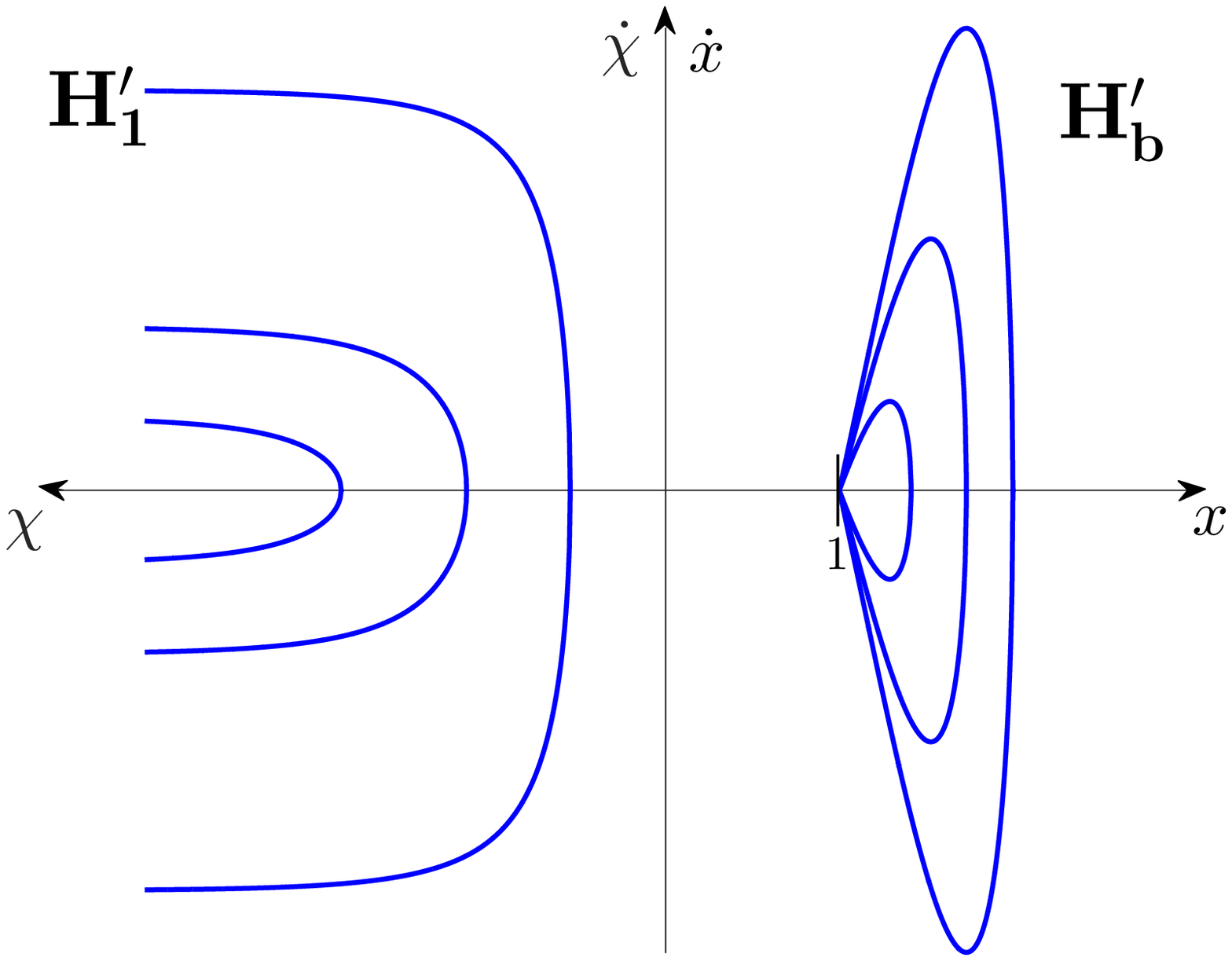}
\caption{ Phase portraits of
{\bf  H${'}_\text{1}$} and {\bf  H${'}_\text{b}$} systems. }
\label{Hb1}
\end{center}
\end{figure}

The phase space portraits in coordinates $(x,\dot{x})$ for the
 {\bf H}-family 
of the systems described by Lagrangians
of the form  $L=\frac{1}{4}m(x)\dot{x}^2 -u_1(x)$ are
presented in Figures  \ref{HaH1}, \ref{HbHc}, \ref{Hb1}. 
 In the case {\bf  H${}_\text{a}$},
as in the case {\bf  H${}_\text{1}$} with $m=1$, coordinate $x$ 
can vary 
on all the real line, and  trajectories in these two cases
have a similar nature: they are bounded  for energies 
$-1\leq E<0$, and unbounded for $E\geq 0$. 
It is interesting to note that 
the 
peculiarity of the case 
{\bf  H${}_\text{a}$} is
that all the phase space trajectories in it are 
conical sections. Namely, 
for $-1<E<0$ these are ellipses,
$\frac{\dot{x}^2}{a^2}+\frac{x^2}{b^2}=1$
with $a^2=4(1+E)$, $b^2=(1+E)/(-E)$,
which degenerate into a point $x=\dot{x}=0$ at $E=-1$.
The case $E=0$ corresponds to sepatrices which here 
 are straight lines $\dot{x}=\pm 2$,
while for $E>0$ the trajectories are  hyperbolas
$\frac{\dot{x}^2}{a^2}-\frac{x^2}{b^2}=1$ with
$a^2=4(1+E)$, $b^2=(1+E)/E$.

In the cases {\bf  H${}_\text{b}$}, {\bf  H${}_\text{d}$}
and {\bf  H${}_\text{e}$} the variable $x$ varies in 
finite intervals, and phase space portraits 
in these cases have a similar nature.
For  $-1\leq E<0$ the trajectories are smooth 
curves lying between the extrema $x_1$ and $x_2$
of the corresponding intervals shown in Table
\ref{Table1},  with returning points 
$x_+=-x_-$, $x_1< x_- < x_+  < x_2$.
For $E=0$ separatrices have cusps
at $x_1$ and $x_2$, which reflect the fact that though
the time necessary to arrive at these points is infinite,
the derivative $d\dot{x}/dx$ at these points
turns into zero. For $E>1$, the slopes 
 $d\dot{x}/dx$ are finite at $x_{1,2}$
 (and time to arrive at these points is infinite).
 The limiting points $x_{1,2}$ and the infinity of time 
necessary  to arrive at them for trajectories with 
 $E\geq 0$ are associated with 
 poles of the position-dependent mass.

In the case {\bf  H${}_\text{c}$},
trajectories are bounded for $-1\leq E<0$, with returning points $x_{\pm}$, 
$0<x_-<x_+<\infty$. For $E=0$, separatrix has a  cusp at $x=0$, 
and asymptotes $\dot{x}=\pm 4$ for $x\rightarrow +\infty$. 
For $E>0$, the trajectories are unbounded, with asymptotes given by
$\frac{d\dot{x}}{dx}=\pm 2\sqrt{E}$ for $x\rightarrow +\infty$.

In the case  {\bf  H${}_\text{b}$}, 
one can consider the infinite domains 
$x>1$ or $x<-1$ instead of the finite interval $x\in(-1,1)$, 
where the mass function also takes positive values. 
Let $x>1$, and denote this case as 
{\bf  H}${}^{\bf \prime}{}_\text{\bf b}$.
Then $x=x(\chi)=-\coth \chi$, $\chi\in(-\infty,0)$, $x\in(1,\infty)$. 
In terms of $\chi$ (after the change of variable),
we have a singular potential $U_1(\chi)=\frac{1}{\sinh^2\chi}$,
and phase space trajectories with $E>0$ are unbounded, 
see Figure \ref{Hb1}. The returning point $\chi_0$ is given by 
 $\sinh\chi_0=-1/\sqrt{E}$, 
and  asymptotes are $\dot{\chi}_\pm=\pm 2\sqrt{E}$.
In coordinates ($x$, $\dot{x}$), however, trajectories 
are confined in the  region $1<x\leq x_0$,
$x_0^2=E+1$, where the returning point $x_0$ corresponds to 
the returning point
$\chi_0$, while $x=1$ corresponds to the asymptotes  
with $\chi\rightarrow -\infty$,
where we have  $\frac{d \dot{x}}{dx}\vert_{x=1}=\pm 2\sqrt{E}$.

\vskip1cm

For position-dependent mass function 
we have $m(x)\rightarrow 0$ for $x\rightarrow x_{1,2}$ and 
unique maximum $m(0)=1$ in the case 
{\bf  H${}_\text{a}$}. On the other hand, 
$m(x)\rightarrow \infty$ for $x\rightarrow x_{1,2}$
in the cases {\bf  H${}_\text{b}$}, {\bf  H${}_\text{d}$}
and {\bf  H${}_\text{e}$}, with the unique minimum $m(0)=1$. 
In the case {\bf  H${}_\text{c}$}, the 
mass function changes monotonically:
$m(x)\rightarrow \infty$ for $x\rightarrow x_{1}=0$ and 
$m(x)\rightarrow 0$ for $x\rightarrow x_{2}=\infty$.

%%%%%%%%%%%%%%%%%%

%%%%%%%%%%%%%%%%%%%%%%%%%%%%
 %%%%%%%%%%%%%%%%%%%%%%%%%%%%%%%%%
  
%%%%%%%%%%%%%%%%%%%%%%%%%%%
%%%%%%%%%%%%%%%%%%%%%%%%%%%%%%%%
\subsection{ 
Trigonometric family
}

Consider now the trigonometric {\bf T}-family of the systems.
Similarly to the {\bf H}-family, 
 Lagrangians for the systems presented in Table \ref{Table2},
$L=\frac{1}{4}m(x)\dot{x}^2-C_n u_1(x)$, 
can be obtained by starting from  
a particle with position-independent mass 
in two-dimensional Euclidean space  and subjected to the action 
of repulsive  Calogero potential, 
$L=\frac{1}{4}(\dot{X}^2+\dot{Y}^2) -C_n\frac{1}{Y^2}$,
and then restricting the motion  
to the  semicircle  $X^2+Y^2=1$, $Y>0$.  
Six different parametrizations of  the semicircle 
$Y=\sqrt{1-X^2}$
given by the functions $X(x)$ and $Y(x)$  
presented in Table \ref{Table2}
result in six models for finite-gap  systems 
shown there. The mass function can be presented here
as $m(x)=X'^2/(1-X^2)$, or $m(x)=Y'^2/(1-Y^2)$.

\begin{table}[h]
\caption{\underline{{\bf T}-family.}\\
Corresponding  finite-gap systems 
are given by potentials 
$u_{{}_{C_n}}(x)=C_n u_1(x)$ with $C_n=n(n+1)\hbar^2$, $n=1,2,\ldots$.
Here $\chi\in(-\frac{\pi}{2},\frac{\pi}{2})$, 
$\text{gd}^{-1}\,x=\text{arctanh}\,(\sin x)$ is the  inverse
gudermannian function
\cite{NIST}. 
% \textcolor{blue}{http://dlmf.nist.gov/4.23#viii},
}
\label{Table2}
\begin{center}
%\begin{tabular}{|c||c|c|c|c|c|c|c|c|} \hline 
\begin{tabular}{|c||c|c|c|c|cV{2.5}c|c|c|c|} \hline 
\multirow{2}{2em}{Case}
 &\multirow{2}{2em}{$m(x)$} & \multirow{2}{2em}{$\varphi(x)$} & \multirow{2}{4em}{$x=x(\chi)$} 
 & \multirow{2}{3em}{$(x_1,x_2)$} & \multirow{2}{2em}{$u_1(x)$} & \multirow{2}{2em}{$X(x)$} & \multirow{2}{2em}{$Y(x)$} \\
 \, & \, & \, & \, & \, & \, & \, &  \\
%\toprule 
\hline
\hline
%%%
%\hskip-2cm
{\small  {\bf \, T${}_\text{1}$} }\, & 
{\small  $1$} & {\small  $1$}  & {\small  $\chi$} & 
{\small  $(-\frac{\pi}{2},\frac{\pi}{2})$} & {\small $\frac{1}{\cos^2 x}$} & {\small $\sin x$}&{\small $\cos x$} 
 \\[3pt]\hline
%%%
%\hskip-2cm
{\small  {\bf \,  T${}_\text{a}$} }\, & 
{\small $\frac{1}{1- x^2}$} & {\small $\cos  \chi$}  &  
{\small $\sin\chi$} &  {\small (-$1,1$)} & 
{\small $\frac{1}{1-x^2}$} & {\small $x$}&{\small $\sqrt{1-x^2}$} 
 \\[3pt]
 \hline
%\hskip-1.9cm
{\small  {\bf \, T${}_\text{b}$} }\, & 
{\small $\frac{1}{(1+x^2)^2}$} & 
{\small $\frac{1}{\cos^2\chi}$}  & 
{\small  $\tan\chi$} & {\small  $(-\infty,\infty)$} & {\small $x^2+1$} & {\small $\frac{x}{\sqrt{1+x^2}}$}&{\small$\frac{1}{\sqrt{1+ x^2}}$} 
 \\[3pt]\hline
 %\hskip-1.8cm
{\small  {\bf \, T${}_\text{c}$} }\, & 
{\small $\frac{1}{e^x-1}$} & {\small $\frac{1+\sin\chi}{\cos\chi}$}  &  
{\small $\ln\frac{2}{1-\sin\chi}$} & {\small  $(0,\infty)$} & 
{\small $\frac{1}{4}\frac{e^{2x}}{e^x-1}$} & 
{\small $1-2e^{-x}$}&
{\small $2\sqrt{e^{-x}-e^{-2x}}$} 
 \\[3pt]\hline
%\hskip-1.8cm
{\small  {\bf \, T${}_\text{d}$} }\, & 
{\small $\frac{1}{\cosh^2x}$} & 
{\small $\frac{1}{\cos \chi}$}   &  
{\small $\text{gd}^{-1}\,\chi$} & {\small  $(-\infty,\infty)$}     & {\small  $\cosh^2 x$} &
 {\small $\tanh x$}&{\small  $\frac{1}{\cosh x}$}  
  \\[3pt]\hline
%%%
%\hskip-1.8cm
{\small  {\bf \, T${}_\text{e}$} }\, &
{\small $\text{dn}^2\, x$} & {\small $\frac{1}{\sqrt{1-k^2\sin^2\chi}}$}  & 
{\small $\text{sn}^{-1}(\sin \chi)$} & {\small $(-\text{\bf K},\text{\bf K})$} &
{\small $\text{nc}^2\, x$} & {\small $\text{sn}\, x$}&{\small   $\text{cn}\,x$}  
\\[5pt]
%%%%
\hline 
\end{tabular}

\end{center}
\end{table}

%%%%%%%%%%%%%%%%%%%%%%%%%%%
%%%%%%%%%%%%%%%%%%%%%%%%%%%%%%%%

As in the case of the {\bf H}-family, 
the kinetic term for {\bf T}-models can also be obtained 
by restricting the kinetic term of a particle on the Riemann sphere to some of 
its 
geodesics.  Take the  metric on the  Riemann sphere in the form
 $ds^2=\frac{dX^2+dY^2}{(1+X^2+Y^2)^2}$, $-\infty <X,Y<\infty$.
Restricting the kinetic term $v^2\equiv (ds/dt)^2$ 
to the geodesic $X^2+Y^2=1$ parametrized by 
$X=\sin x$, $Y=\cos x$, we reproduce the kinetic term for 
{\bf T${}_\text{1}$} case with $m=1$.
Restriction of $\frac{1}{4}v^2$ to the geodesic $Y=0$ with 
subsequent change of the notation $X\rightarrow x$ 
 yields  the kinetic term for {\bf T${}_\text{b}$} model.
By appropriate change of the variable $x$, which 
can be found from the column $X(x)$ of the Table,  
one can reproduce all other
kinetic terms for {\bf T}-models. 

The case {\bf T${}_\text{e}$} in the limits $k\rightarrow 0$ and 
$k\rightarrow 1$  transforms into
 the {\bf  T${}_\text{1}$} and  {\bf  T${}_\text{d}$} cases, 
 respectively.
The family  {\bf T${}_\text{e}$}  can be considered therefore as that 
interpolating continuously between the 
position-independent mass,
{\bf  T${}_\text{1}$}, and PDM,
{\bf  T${}_\text{d}$}, cases
of trigonometric finite-gap systems.

After the application of similarity transformation 
and the change of variable, we reduce all the cases to the 
corresponding quantum systems from the case
{\bf T${}_\text{1}$}. Such a system
characterized 
by the integer parameter $n$   can be obtained by 
subsequent application of $n$ Darboux transformations 
to the free particle ($n=0$) confined into the infinite potential well
with impenetrable walls at $\chi_1=-\frac{\pi}{2}$ and 
$\chi_2=\frac{\pi}{2}$ \cite{SUSYQM}. 
Energy levels of the bound states are $(n+l+1)^2-1$, $l=0,1,\ldots$.
Though the Lax-Novikov integral can formally be obtained from hyperbolic case 
by the transformation $\chi\rightarrow i\chi$, 
this is a non-physical operator:  its action on the physical states
produces the states divergent at the edges $\chi_{1,2}=\pm \frac{\pi}{2}$ of the 
interval. The model {\bf  T${}_\text{1}$}
is often called in the literature the Higgs 
oscillator~\cite{Higgs,Kalnins,Armen,Balles,Evnin}.

\begin{figure}[htbp]
\begin{center}
\includegraphics[scale=0.44]{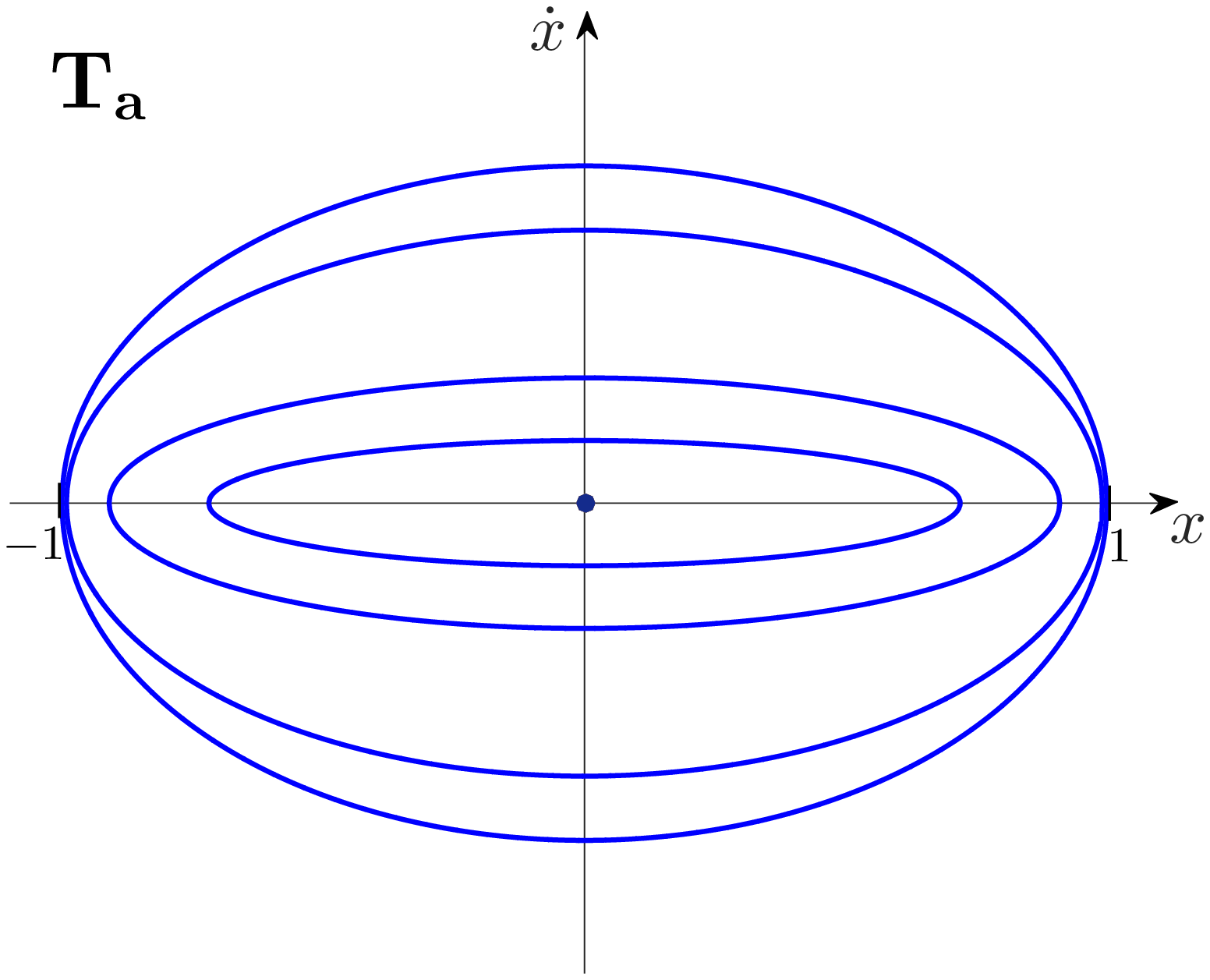}\includegraphics[scale=0.44]{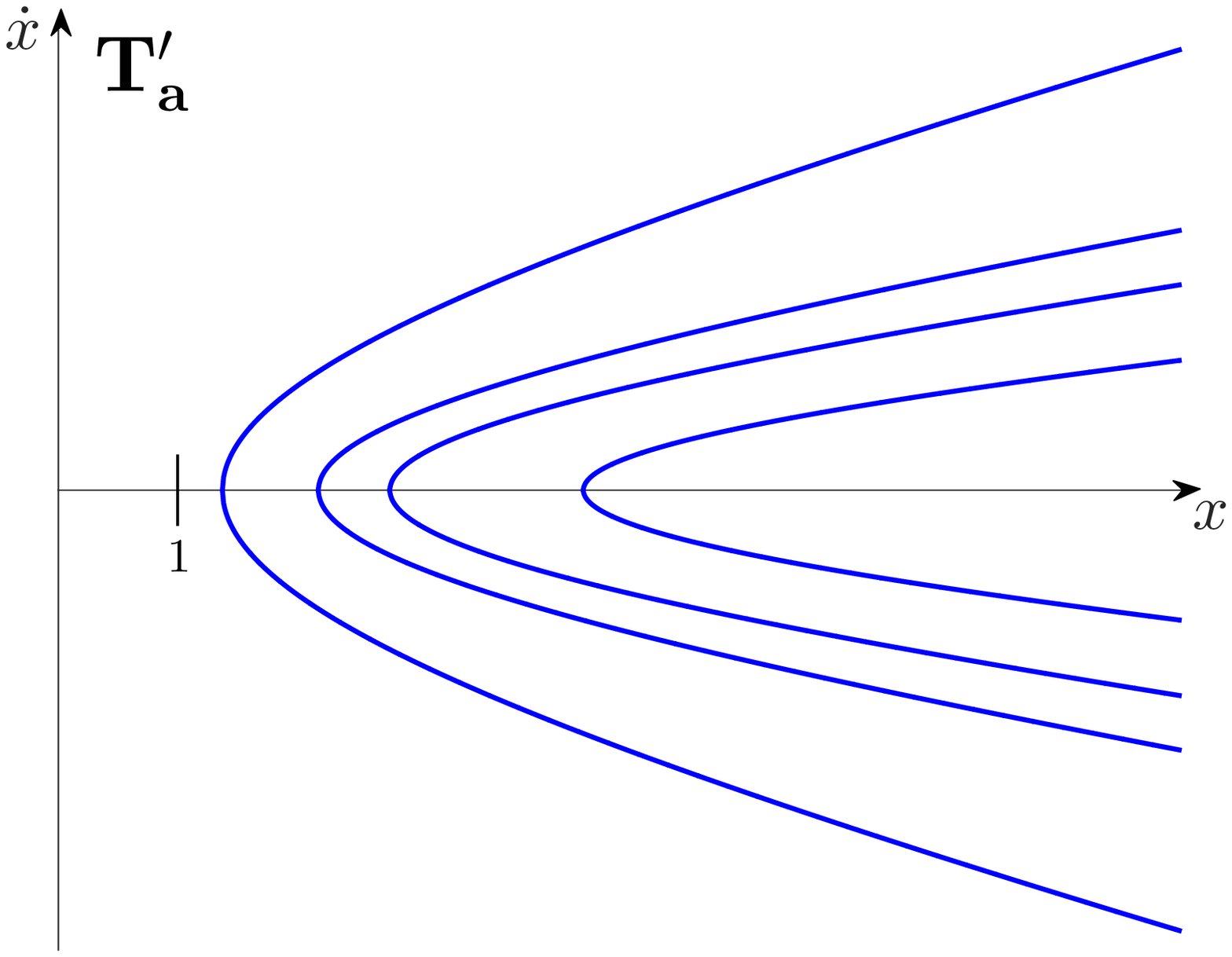}
\caption{
Phase portraits of
{\bf  T}${}_\text{\bf a}$ and {\bf T${'}_\text{a}$} systems. }
\label{TaTa1}
\end{center}
\end{figure}

The phase portraits for 
{\bf  T${}_\text{1}$}, {\bf  T${}_\text{a}$}
and {\bf  T${}_\text{e}$} cases are similar.
All the trajectories with $E\geq 1$ in these cases,
where $E=1$ corresponds to the minimum value of the 
potentials $u_1$,  are bounded and  closed,
with returining points  $x_-=-x_+$,  $x_1<x_-<x_+<x_2$.
There is, however, a  difference between these cases:
 in the  {\bf  T${}_\text{1}$} case, the extrema points $x_{1,2}$ of the domain 
 of $x$ correspond to singular points of the
potential, while for {\bf  T${}_\text{a}$}
and {\bf  T${}_\text{e}$} cases 
 they correspond to  zeros 
of the $1/m(x)$ function. 
The peculiarity of the {\bf  T${}_\text{a}$} case is also that all
the trajectories in it are ellipses: 
$\frac{\dot{x}^2}{a^2}+\frac{x^2}{b^2}=1$ with 
$a^2=4(E-1)$, $b^2=(E-1)/E$.
This phase portrait is  similar to that of the  harmonic oscillator,
with the difference that  here $b^2$ is restricted from above, 
$b^2<1$, see Figure \ref{TaTa1}.

\begin{figure}[htbp]
\begin{center}
\includegraphics[scale=0.44]{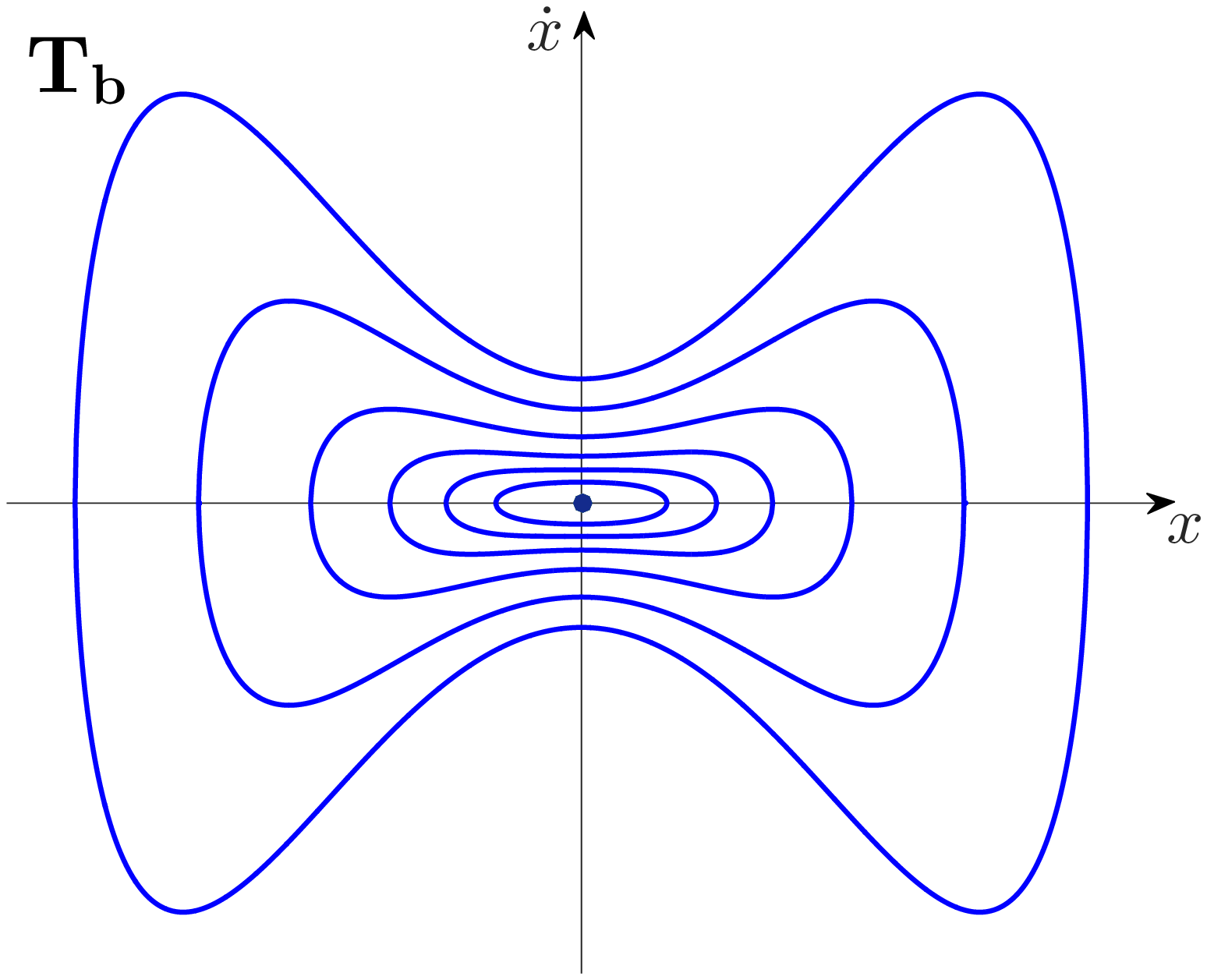}\includegraphics[scale=0.44]{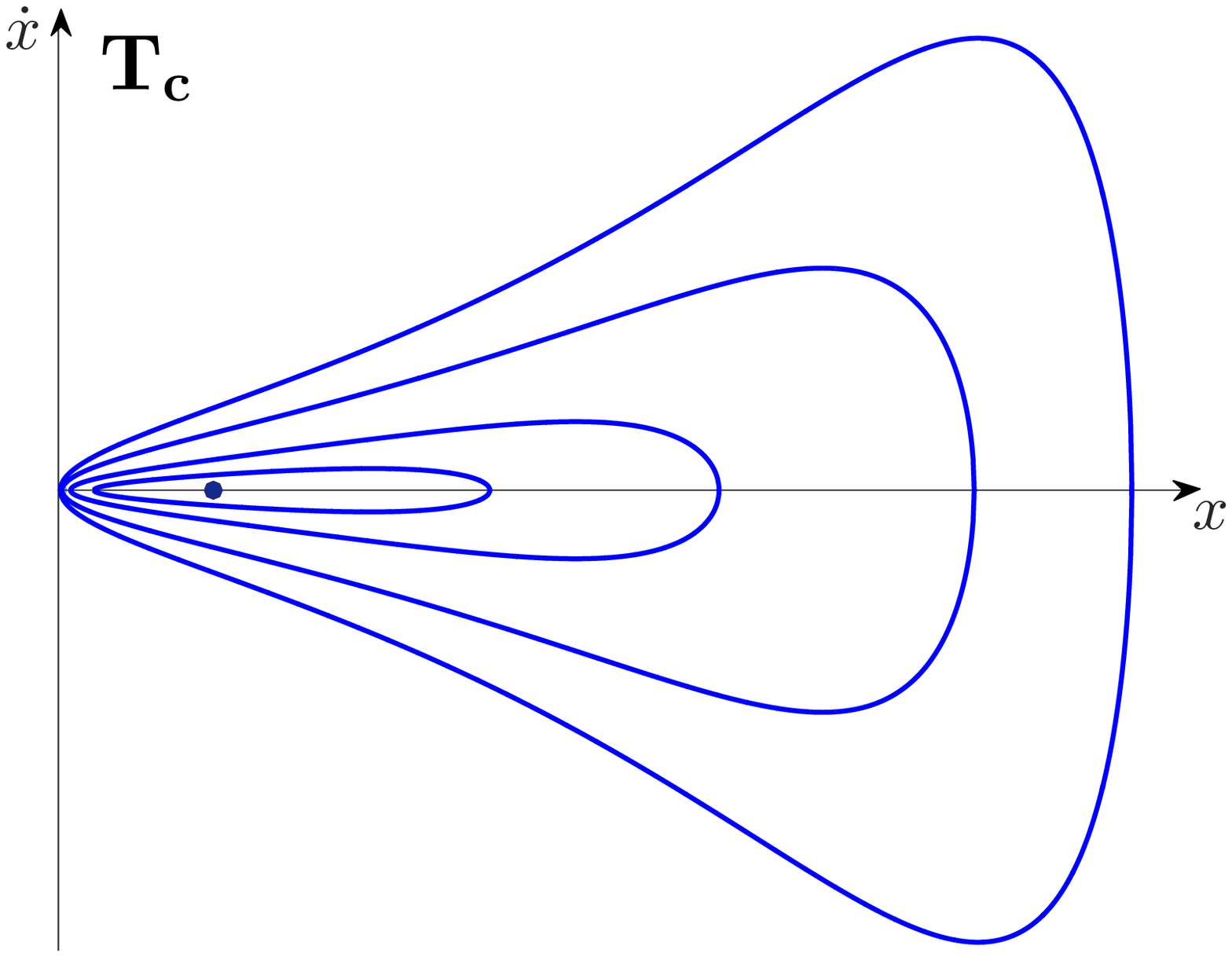}
\caption{
Phase portraits of
{\bf  T}${}_\text{\bf b}$ and {\bf T}${}_\text{\bf c}$ systems.
}
\label{TbTc}
\end{center}
\end{figure}

In the case of {\bf  T${}_\text{c}$},  all the 
trajectories are closed, with returning points $x_-$ and $x_+$ satisfying the relation
$0<x_-<x_0<x_+<\infty$, where $x_0=\ln 2$ corresponds to the 
point  where potential takes the minimum value
$1$, i.e., unlike the three above mentioned cases, 
here the values of the returning point 
$x_+$ are not bounded, see Figure \ref{TbTc}.
 The phase portraits in $(x,\dot{x})$ coordinates for the 
cases {\bf  T${}_\text{b}$} (note that in this case 
potential is quadratic) 
and {\bf  T${}_\text{d}$}
are similar to the phase portrait for a usual one-dimensional
harmonic oscillator: the trajectories are closed smooth curves 
whose sizes increase with increasing of energy $E=\frac{1}{4}m(x)\dot{x}^2
+u_1(x)$.
There is, however, a difference
 in comparison with the harmonic oscillator.
 As it is seen from Figure \ref{TbTc}, the trajectories in the  {\bf  T${}_\text{b}$} case 
 are convex 
 only for $1\leq E\leq E_*=\frac{3}{2}$, 
 but they loose this property for $E>E_*$.
In the case  {\bf  T${}_\text{d}$}, $E_*=2$. 
Similar properties related to (non-)convexitivity
of phase space trajectories 
is also characteristic of {\bf  T${}_\text{c}$}
case.

The model {\bf  T${}_\text{a}$} 
can be modified for the  case  {\bf  T}${\bf {}'}_\text{\bf a}$ 
with  $x^2>1$  
by multiplying  Lagrangian
by $-1$ to have a positive-valued mass function,
$L=\frac{1}{4}\tilde{m}(x)\dot{x}^2-\tilde{u}_1(x)$,
where $\tilde{m}(x)=\frac{1}{x^2-1}$, $\tilde{u}_1(x)=\frac{1}{x^2-1}$.
In this case 
$x(\chi)=\cosh \chi$ with $\chi\in(0,\infty)$  and  
$x\in(1,\infty)$, 
 and after the change of variable
we obtain a singular Lagrangian $\tilde{U}_1(\chi)=\frac{1}{\sinh^2\chi}$
exactly as in the case  {\bf  T}${\bf {}'}_\text{\bf b}$.
So, in coordinates $\chi$, $\dot{\chi}$ the unbounded
trajectories for $E>0$ are exactly of the same form
described above for 
the singular finite-gap model  {\bf  H}${\bf {}'}_\text{\bf 1}$
but with $\chi<-1$ there changed for $\chi>1$ 
in the case  {\bf  T}${\bf {}'}_\text{\bf 1}$ (we do not
show these trajectories in coordinates $(\chi,\dot{\chi})$ here).
Unlike the  {\bf  H}${\bf {}'}_\text{\bf b}$ model, 
 here, in the {\bf  T}${\bf {}'}_\text{\bf a}$ model, 
the trajectories are the hyperbolas
$\frac{\dot{x}^2}{a^2}-\frac{x^2}{b^2}=-1$ 
with $a^2=4(1+E)$, $b^2=(1+E)/E$, $E>0$, see 
Figure \ref{TaTa1}.

In the case {\bf  T${}_\text{a}$}, function $m(x)$ tends to infinity 
when $x\rightarrow x_{1,2}$, taking minimum value $m(0)=1$.
In the cases {\bf  T${}_\text{b}$} and {\bf  T${}_\text{d}$},
$m(x)\rightarrow 0$ when $x\rightarrow x_{1,2}$,
and takes maximum value $m(0)=1$.
In the case {\bf  T${}_\text{e}$},  $m(x)\rightarrow k'^2>0$
for $x\rightarrow x_{1,2}$, and takes maximum value $m(0)=1$.
In the case {\bf  T${}_\text{c}$}, $m(x)$ changes monotinically,
with $m(x)\rightarrow \infty$ when $x\rightarrow 0$ and 
$m(x)\rightarrow 0$ as $x\rightarrow \infty$.

%%%%%%%%%%%%%%%%%%%%%%%%%%%%%
 %%%%%%%%%%%%%%%%%%%%%%%%%%%%%%%%%%

 \subsection{Finite-gap elliptic {\bf L}-- and {\bf D}--families}

\begin{figure}[htbp]
\begin{center}
\includegraphics[scale=0.5]{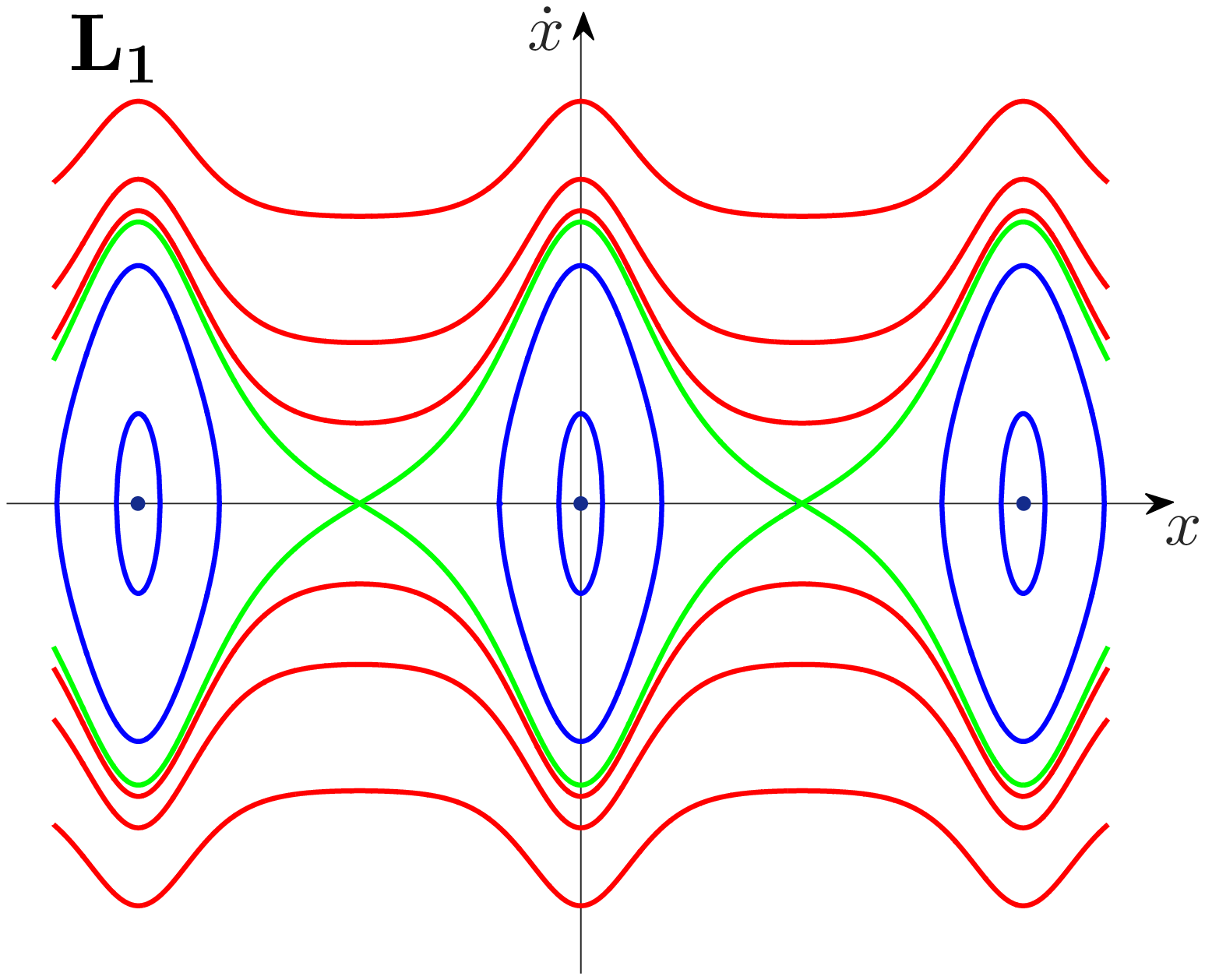}
\caption{Phase portrait of 
{\bf  L${}_\text{1}$} system with $k=0.99$.}
\label{L1Fig}
\end{center}
\end{figure}

Consider now finite-gap elliptic generalizations of the  hyperbolic and 
trigonometric families which are presented in  Table \ref{Table3}.
The phase portrait of the  case 
{\bf L${}_\text{1}$}, which is a periodic generalization of the 
{\bf H${}_\text{1}$} case,  is shown in Figure \ref{L1Fig}.
 
%\newpage

\begin{table}[htbp]
\caption{\underline{Elliptic {\bf L}-- and {\bf D}--families.}\\
Here 
 $x_{{}_E}(\chi)=\frac{1}{k'}
 \ln\left(1+k'\frac{\text{dn}\,\chi+k'\text{sn}\,\chi}{1-\text{sn}\,\chi}\right)$,
 $u_{{}_{L_{E}}}(x)=-k'^2\left(
 \frac{e^\xi-k^2e^{-\xi}}{e^\xi+k^2e^{-\xi}-2k^2}\right)^2$, 
 $u_{{}_{D_{E}}}(x)=\frac{1}{4}\frac{(
 e^\xi-k^2e^{-\xi})^2}{e^\xi+k^2e^{-\xi}-1-k^2}$, $\xi=k'x$;
 $\text{am}\,\chi=\text{am}\,(\chi,k)=\text{Arcsin}\,\sn\,(\chi,k)$ is Jacobi's 
 amplitude function \cite{WW,NIST}.
 The limiting cases  {\bf H}${}_{\alpha\, {\bf \pm}}$ are defined in the same way as 
   {\bf H}${}_{\alpha}$, $\alpha=${\bf 1},{\bf a},{\bf b},{\bf c},{\bf d}, but with 
$u_1(x)$ changed for 
 $u_\pm(x)=\pm 1$.
 Analogously,   {\bf T}${}_{\alpha\, -}$ is defined 
 as {\bf T}${}_{\alpha}$  but with  potential 
 $u_1(x)$ changed for 
 $u_-(x)=-1$.
 }
\label{Table3}
\begin{center}
\begin{tabular}{|c||c|c|c|c|c|cV{2.5}c|c|c|c|} \hline
\multirow{2}{1.6em}{\small Case}
&\multirow{2}{1.8em}{$m(x)$}& \multirow{2}{1.7em}{$\varphi(\chi)$} & \multicolumn{2}{c|}{$x = x(\chi)$} &
 \multirow{2}{2em}{$u_1(x)$} & \multirow{2}{2.3em}{$U_1(\chi)$}  & \multirow{2}{2.7em}{$k\rightarrow 1$} &
  \multirow{2}{2.7em}{$k\rightarrow 0$}\\
\cline{4-5}
 &\, & \,  &{\small $(\chi_1,\chi_2)$} & {\small $(x_1,x_2)$}  & \, &\, &\, &\,   \\
%\toprule  
\hline
\hline
\multirow{3}{1em}{{\small  {\bf L${}_\text{1}$} }}\, & \multirow{4}{1em}{1} &
\multirow{4}{1em}{1} & \multicolumn{2}{c|}{$\chi$}& \multirow{2}{3em}{{\small $-\text{dn}^2\,x$}} & 
\multirow{2}{2.5em}{{\small $-\text{dn}^2\, \chi$}}& \multirow{3}{1em}{{\small  {\bf H${}_\text{1}$} }} &
\multirow{3}{2em}{{\small  {\bf  H${}_\text{1\,--}$} }}\\
\cline{4-5}
& \,  & \, & {\small  $(-\infty,\infty)$} & {\small  $(-\infty,\infty)$}  & \, &\, &\,&\, \\
\cline{1-1} \cline{4-9}
\multirow{3}{1em}{{\small  {\bf D${}_\text{1}$} }}\, & \, &\, &  {\small (-$\text{\bf K}$, $\text{\bf K}$)} & 
 {\small (-$\text{\bf K}$, $\text{\bf K}$)}
&\multirow{2}{2em}{{\small $\text{dc}^2\,x$}} & \multirow{2}{2em}{{\small $\text{dc}^2\,\chi$}} &
\multirow{3}{1em}{{\small  {\bf  H${}_\text{1+}$}} } &
\multirow{2}{1.8em}{{\small  {\bf T${}_\text{1}$} }}\\
\cline{4-5}
& \,  & \,   & \multicolumn{2}{c|}{\,}  & \, &\, &\,&\, \\
\Xhline{0.9pt}
\multirow{3}{1em}{{\small  {\bf L${}_\text{A}$} }}\, & \multirow{4}{5.3em}{\small $\frac{1}{(1-x^2)(1-k^2x^2)}$} &
\multirow{4}{4em}{\small $\text{cn}\,\chi\,\text{dn}\,\chi $}  & 
\multicolumn{2}{c|}{\,}& 
\multirow{2}{4em}{{\small $k^2x^2-1$ }} & 
\multirow{3}{3em}{{\small $-\text{dn}^2\,\chi$ }}&
\multirow{3}{1em}{{\small  {\bf H${}_\text{b}$} }} &
\multirow{2}{2em}{{\small  {\bf  T${}_\text{a\,--}$} }}\\
%\cline{4-5}
& \,  & \, & \multicolumn{2}{c|}{\small  $\text{sn}\,\chi$}  & \, &\, &\,&\, \\
\cline{1-1} \cline{4-9}
\multirow{3}{1em}{{\small  {\bf D${}_\text{A}$} }}\, &
 \, & \,& {\small (-$\text{\bf K}$, $\text{\bf K}$)} 
 & {\small  $(-1,1)$} 
 &  \multirow{2}{4em}{{\small $\frac{1-k^2x^2}{1-x^2}$}} & 
 \multirow{3}{2em}{{\small $\text{dc}^2\,\chi$ }}
&\multirow{3}{1em}{{\small  {\bf  H${}_\text{b+}$} }} &\multirow{2}{2em}{{\small  {\bf T${}_\text{a}$} }}\\
\cline{4-5}
& \,  & \,   & \multicolumn{2}{c|}{\,}  & \, &\, &\,&\, \\
\Xhline{0.9pt}
\multirow{3}{1em}{{\small  {\bf L${}_\text{B}$} }}\, & \multirow{4}{5.3em}{\small $\frac{1}{(1+x^2)(1+k'^2x^2)}$} &
\multirow{4}{4em}{\small $\text{dn}\,\chi\,\text{nc}^2\,\chi $}  &
 \multicolumn{2}{c|}{\,}& 
\multirow{2}{4em}{{\small $-\frac{1+k'^2x^2}{1+x^2}$}} & 
\multirow{3}{3em}{{\small $-\text{dn}^2\,\chi$ }}&
\multirow{3}{1em}{{\small  {\bf H${}_\text{a}$} }} &
\multirow{3}{2em}{{\small  {\bf  T${}_\text{b\,--}$} }}\\
%\cline{4-5}
& \,  & \, & \multicolumn{2}{c|}{\small  $\text{sc}\,\chi$}  & \, &\, &\,&\, \\
\cline{1-1} \cline{4-9}
\multirow{3}{1em}{{\small  {\bf D${}_\text{B}$} }}\, &
 \, & \,& {\small (-$\text{\bf K}$, $\text{\bf K}$)} 
 & {\small  $(-\infty,\infty)$} 
 &  \multirow{2}{4em}{{\small $k'^2 x^2+1$}} & 
 \multirow{3}{2em}{{\small $\text{dc}^2\,\chi$ }}
&\multirow{3}{1em}{{\small  {\bf  H${}_\text{a+}$} }} &\multirow{2}{2em}{{\small  {\bf T${}_\text{b}$} }}\\
\cline{4-5}
& \,  & \,   & \multicolumn{2}{c|}{\,}  & \, &\, &\,&\, \\
\Xhline{0.9pt}
\multirow{3}{1em}{{\small  {\bf L${}_\text{C}$} }}\, & \multirow{4}{4em}{\small $\frac{1}{1-k^2\sin^2x}$}&
\multirow{4}{2em}{\small $\text{dn}\,\chi$}   & \multicolumn{2}{c|}{\small  $\text{am}\,\chi$}&
\multirow{2}{5em}{{\small $k^2\sin^2 x-1$}} & \multirow{3}{3em}{{\small $-\text{dn}^2\,\chi$ }}
&\multirow{3}{1em}{{\small  {\bf H${}_\text{d}$} }} &\multirow{3}{2em}{{\small  {\bf  H${}_\text{1\,--}$} }}\\
\cline{4-5}
& \,  & \, & {\small  $(-\infty,\infty)$} & {\small  $(-\infty,\infty)$}  & \, &\, &\,&\, \\
\cline{1-1} \cline{4-9}
 \multirow{3}{1em}{{\small  {\bf D${}_\text{C}$} }}\, & 
 \, & \,& \, {\small (-$\text{\bf K}$, $\text{\bf K}$)} 
 & {\small (-$\frac{\pi}{2}$, $\frac{\pi}{2}$)} &
  \multirow{2}{4em}{{\small $\frac{1-k^2\sin^2x}{\cos^2 x}$}} & \multirow{3}{2em}{{\small $\text{dc}^2\,\chi$ }}
&\multirow{3}{1em}{{\small  {\bf  H${}_\text{d+}$} }} &\multirow{2}{2em}{{\small  {\bf T${}_\text{1}$} }}\\
\cline{4-5}
& \,  & \,   & \multicolumn{2}{c|}{\,}  & \, &\, &\,&\, \\
\Xhline{0.9pt}
\multirow{3}{1em}{{\small  {\bf L${}_\text{D}$} }}\,  & \multirow{4}{4em}{\small $\frac{1}{1+k'^2\sinh^2x}$}&
\multirow{4}{2em}{\small $\text{dc}\,\chi$}   & \multicolumn{2}{c|}{\,} &\multirow{2}{5.6em}{{\small $-k'^2-\frac{k^2}{\cosh^2 x}$}} 
& \multirow{3}{3em}{{\small $-\text{dn}^2\,\chi$ }}
&{\multirow{3}{1em}{\small  {\bf H${}_\text{1}$} }} &\multirow{3}{2em}{{\small  {\bf T${}_\text{d\,--}$} }}\\
%\cline{4-5}
& \,  & \, & \multicolumn{2}{c|}{\small  $\text{arcsinh}\,(\text{sc}\,\chi)$}  & \, &\, &\,&\, \\
\cline{1-1} \cline{4-9}
 \multirow{3}{1em}{{\small  {\bf D${}_\text{D}$} }}\, & \, & \,& {\small (-$\text{\bf K}$, $\text{\bf K}$)} &
  {\small  $(-\infty,\infty)$} & \multirow{2}{5.5em}{{\small $k'^2\sinh^2 x+1 $}} & \multirow{3}{2em}{{\small $\text{dc}^2\,\chi$ }}
&\multirow{3}{1em}{{\small  {\bf  H${}_\text{1+}$} }} &\multirow{2}{2em}{{\small  {\bf T${}_\text{d}$} }}\\
\cline{4-5}
& \,  & \,   & \multicolumn{2}{c|}{\,}  & \, &\, &\,&\, \\
\Xhline{0.9pt}
\multirow{3}{1em}{{\small  {\bf L${}_\text{E}$} }}\,& \multirow{4}{5.1em}{\small $\frac{k'^2}{(1-e^{-\xi})(e^\xi-k^2)}$} &
\multirow{4}{4em}{\small $\text{nc}\,\chi + \text{sc}\,\chi$}  & \multicolumn{2}{c|}{\,} &
\multirow{2}{3em}{{\small $u_{{}_{L_{E}}}(x)$}} & \multirow{3}{3em}{{\small $-\text{dn}^2\,\chi$ }}
&\multirow{3}{1em}{{\small  {\bf H${}_\text{c}$} }} &
\multirow{2}{2em}{{\small  {\bf  T${}_\text{c\,--}$} }}\\
%\cline{4-5}
& \,  & \, & \multicolumn{2}{c|}{\small  $x_{{}_E}(\chi)$ }  & \, &\, &\,&\, \\
\cline{1-1} \cline{4-9}
\multirow{3}{1em}{{\small  {\bf D${}_\text{E}$} }}\,& \, & \,&
 {\small (-$\text{\bf K}$, $\text{\bf K}$)} & {\small  $(0,\infty)$} &
\multirow{2}{3em}{{\small $u_{{}_{D_{E}}}(x)$}} & \multirow{3}{2em}{{\small $\text{dc}^2\,\chi$ }}
&\multirow{3}{1em}{{\small  {\bf  H${}_\text{c+}$}} } &\multirow{2}{2em}{{\small  {\bf T${}_\text{c}$} }}\\
\cline{4-5}
& \,  & \,   & \multicolumn{2}{c|}{\,}  & \, &\, &\,&\, \\
\Xhline{0.9pt}

\end{tabular}

\end{center}
\end{table}

 The {\bf L${}_\text{E}$} and {\bf D${}_\text{E}$} cases
can be considered as a generalization
of the cases {\bf H${}_\text{c}$} and {\bf T${}_\text{c}$}. 
To see this  
we note that the function 
$
\varphi_{{}_E}(\chi)=(1+\sn\,\chi)/{\cn\,\chi}
$ 
transforms in the limits $k\rightarrow 1$ and $k\rightarrow 0$ 
into the functions 
$\varphi(\chi)$ of the indicated hyperbolic and trigonometric cases.
By means of (\ref{xchi}) we find  
 $x_{{}_E}(\chi)=\frac{1}{k'}
 \ln\left(1+k'\frac{\text{dn}\,\chi+k'\text{sn}\,\chi}{1-\text{sn}\,\chi}\right),
$ 
and then
$
\sn\,\chi=1-{2k'^2}({e^\xi-2k^2+k^2e^{-\xi}})^{-1},
$ 
$
\xi=k'x\,.
$
This allows us to identify the potentials 
 $u_{{}_{L_{E}}}(x)$ and  $u_{{}_{D_{E}}}(x)$
 using the indentities  
 $\text{dn}^2\chi=1-k^2\text{sn}^2\,\chi$
 and  $\text{cn}^2\,\chi=1-\text{sn}^2\,\chi$.
For  $f(x)=\varphi(\chi)\vert_{\chi=\chi(x)}$,
we obtain 
$
f(x)=\frac{1}{k'}\sqrt{(1-e^{-\xi})(e^\xi-k^2)},
$
 and  finally find the corresponding position dependent mass
 shown in the Table \ref{Table3}. 
The  potentials
$u_{{}_{L_{E}}}(x)$ and $u_{{}_{D_{E}}}(x)$ 
can be presented equivalently as 
\be\label{u(x)perLD+}
u_{{}_{L_{E}}}(x)=-k'^2\frac{\sinh^2 \xi_k}{(\cosh  \xi_k  - k)^2}\,,\qquad
u_{{}_{D_{E}}}(x)=k^2\frac{\sinh^2(\xi_k)}{2k(\cosh \xi_k - k)-k'^2}\,,
\ee
where $\xi_k\equiv \xi-\ln k$. 

Jacobi's amplitude function $\text{am}\,(x,k)$ satisfies relations $\text{am}(x,0)=x$,  
$\text{am}\,(x,1)=\text{gd}\,x$,
and can be considered as a generalization of the 
gudermannian function.
In correspondence with this we note 
that  $x=x(\chi)=\text{am}\,\chi$, which is 
the change of variable function
in the case {\bf L${}_\text{C}$},  
 appears as a generalization 
 of the kink solution in the sine-Gordon model \cite{sGam1,sGam2}. 
 
 In the case {\bf A}, the even mass function $m(x)$ takes minimum value 
 $m(0)=1$ and  $m(x)\rightarrow \infty$ for $x\rightarrow \pm 1$.
 In the cases {\bf B} and {\bf D}, even mass function takes maximum value
 $m(0)=1$ and $m(x)\rightarrow 0$ for $x\rightarrow \pm \infty$.
 In the case {\bf C}, $m(-x)=m(x)$, $m(0)=1$ and $m\rightarrow 1/k'^2$
 for $x\rightarrow \pm \frac{\pi}{2}$.

In correspondence with the behaviour of $U_1(\chi)=\text{dc}^2\,\chi$,
potential $u_1(x)$ in all the cases  of the {\bf D}-family
tends to $+\infty$ when 
$x$ tends to the corresponding edge values $x_{1,2}$,
taking minimum value $+1$ at $x=0$ in all the cases except 
the case  {\bf D${}_\text{E}$}, 
where this happens at $x=\frac{1}{k'}(\ln k
+\text{arccosh}\,\frac{1}{k})$, that is the root
of  the 
 equation $\cosh \xi_k=\frac{1}{k}$.
Analogously, in correspondence with 
$U_1(\chi)=-\text{dn}^2\,\chi$, 
potential $u_1(x)$ in the cases of the {\bf L}-family
tends to the maximum value $-k'^2$ 
when  $x\rightarrow x_{1,2}$, taking minimum value $-1$ 
at $x=0$ in all the cases except 
the case  {\bf L${}_\text{E}$},
where this minimum value is taken at 
$x=\frac{1}{k'}\ln(1+k')$.

{}From the two last columns of the Table \ref{Table3} we also see that elliptic
(after similarity transformation and the change of variable)
 Lam\' e models {\bf L} provide us with some interpolation 
 between reflectionless models {\bf H}
 and corresponding free particle models with 
 position-dependent, or position-independent (unit) mass function.
 Particularly, the case {\bf L${}_\text{B}$}
 provides a finite-gap periodic   generalization of the 
Mathews-Lakshmanan oscillator model described by the case 
 {\bf H${}_\text{a}$},
 while the  {\bf L${}_\text{A}$} case can be considered 
 as a finite-gap periodic generalization of the `inflationary model'
  {\bf H${}_\text{b}$}.
 Analogous job is made by the Darboux-Treibich-Verdier models {\bf D},
 which can be considered as the systems 
 interpolating between the trigonometric models {\bf T}
 and corresponding free particle systems.

We do not discuss the spectrum and corresponding 
Lax-Novikov operators of finite-gap systems 
of the {\bf L}-- and {\bf D}--families here, 
 and just refer 
to  
\cite{CJNP,ACJGP}.

\vskip0.1cm
In conclusion of this subsection,
it is worth to make an additional comment here
to be valid for each of  the three families of finite-gap systems presented above. 
 If after corresponding similarity transformations and changes of variables
 two systems with different position dependent masses $m(x)$ and $\tilde{m}(\xi)$
 and potentials $u(x)$ and $\tilde{u}(\xi)$
produce the same system $H=-\frac{d^2}{d\chi^2}+U(\chi)$, 
the following equality for the quantum kinetic terms has to be valid:
$\sqrt{f}\left(\sqrt{f}\frac{d}{dx}\sqrt{f}\right)^2\frac{1}{\sqrt{f}}\vert_{x=x(\chi)}=
\sqrt{\tilde{f}}\left(\sqrt{\tilde{f}}\frac{d}{d\xi}\sqrt{\tilde{f}}\right)^2\frac{1}{\sqrt{\tilde{f}}}
\vert_{\xi=\xi(\chi)}$.
{}From this equality we find that to establish the relation between the 
quantum kinetic terms of any two
systems presented in the tables which produce the same quantum system  
 $H=-\frac{d^2}{d\chi^2}+U(\chi)$, the following additional 
 similarity transformation 
  is required:
 \be\label{ftildf}
 	\left(\sqrt{f(x)}\frac{d}{dx}\sqrt{f(x)}\right)^2=
\sqrt{\frac{\tilde{f}(\xi)}{f(x)}}\left(\sqrt{\tilde{f}(\xi)}\frac{d}{d\xi}\sqrt{\tilde{f}(\xi)}\right)^2
\sqrt{\frac{f(x)}{\tilde{f}(\xi)}}
\vert_{\xi=\xi(x)}\,.
 \ee
 Here, on the right hand side of Eq. (\ref{ftildf}),
   $\xi=\xi(x)$ is given by $\xi(x)=\xi(\chi)\vert_{\chi=\chi(x)}$, and so, 
 $\tilde{u}(\xi(x))=u(x)$.
 For example,  the system from the elliptic case {\bf A}  can be obtained 
from the corresponding systems of the elliptic case  {\bf C} 
by the changes of variables $x\rightarrow \xi$, $\xi=\sin x$.
In this way the potentials $u(x)$ 
from the case  {\bf C} transform into corresponding potentials 
of the case {\bf A}. 
For  kinetic term  
we have then 
$\left(\sqrt{f_{\text{\bf C}}}\frac{d}{dx}\sqrt{f_{\text{\bf C}}}\right)^2\rightarrow 
(1-x^2)^{1/4}\left(\sqrt{f_{\text{\bf A}}}\frac{d}{dx}\sqrt{f_{\text{\bf A}}}
\right)^2 (1-x^2)^{-1/4}$. Thus after additional similarity  transformation
quantum Hamiltonian $H_{\text{\bf C}}(x)$ transforms into 
$H_{\text{\bf A}}(x)$.

\subsection{Elliptic finite-gap systems and Seiffert's spiral}\label{subSeiffert}
 
 The systems presented in Table \ref{Table3} can be obtained 
 from a particle on the unit sphere 
 subjected to the action of 
 certain potentials of the forms like those 
 indicated at the beginning of 
 the
  section, 
 to which it is necessary to apply 
 a certain reduction procedure.
 To show this we take the $\R^3$  metric 
in cylindrical  coordinates  $ds^2=\rho^2 d\phi^2 +d\rho^2 
+dz^2$. On the surface of the unit sphere 
this can be reduced to one of the two forms 
\be\label{dszrho}
ds^2(z,\phi)=(1-z^2)d\phi^2 +\frac{dz^2}{1-z^2},\qquad
ds^2(\rho,\phi)=\rho^2d\phi^2 +\frac{dz^2}{1-\rho^2}\,,
\ee
where we have 
used the sphere equation $\rho^2+z^2=1$ to eliminate
the dependence on $\rho$ or $z$.
Let us restrict additionally the motion by requiring 
that $d\phi=k ds$, where $k\in\R$ is a constant.
As a result,  
$L_0=\frac{1}{4}(ds/dt)^2$ takes the form 
of the kinetic term for a particle moving  along the Seiffert's spiral
\cite{WW,Seiff}. Particularly, if we take 
$z=\pm \frac{1}{\sqrt{1-x^2}}$, $x\in [-1,1]$,  that corresponds to 
the function 
$Y(x)$ from 
{\bf T${}_\text{b}$} case but with a sign, and use the first  form $ds^2(z,\phi)$  
from (\ref{dszrho}), we reproduce the mass term for 
{\bf L${}_\text{A}$} and {\bf D${}_\text{A}$} cases.
Potential can be chosen initially  in the
Calogero-like  form 
$u_1(z)=-k'^2-\frac{k^2}{z^2}$ for {\bf L} families,
or in the harmonic oscillator form $u_1(z)=k'^2z^2+k^2$
for {\bf D} families. The same {\bf L${}_\text{A}$} and {\bf D${}_\text{A}$}
systems can be reproduced by using $ds^2(\rho,\phi)$,
and setting  $\rho=x$  which corresponds to $X(x)$ 
from the same  {\bf T${}_\text{b}$} 
case~\footnote{In this case $\rho$ with a sign corresponds 
to the horizontal coordinate 
in the meridian  plane $(\rho,z)$, for explanations
see \cite{Seiff}.}. The initial form of potentials 
is interchanged in comparison with the case when we
proceed from the  $ds^2(z,\phi)$   form of the spherical metric:
we should take $u_1(\rho)=-k'^2-\frac{k^2}{\rho^2}$ for the {\bf D} case
and $u_1(\rho)=k'^2\rho^2+k^2$
for the {\bf L} case.
In the same vein one can use other parametrizations for
$z$ and $\rho$ coordinates shown in Table \ref{Table2}
to reproduce the systems  presented 
in  Table \ref{Table3}. The correspondence 
between parametrizations and elliptic families
is the following: 
{\bf T${}_\text{e}$} $\rightarrow$ {\bf 1},
{\bf T${}_\text{b}$} $\rightarrow$ {\bf B},
{\bf T${}_\text{1}$} $\rightarrow$ {\bf C},
{\bf T${}_\text{d}$} $\rightarrow$ {\bf D}.
At the same time, if we take $\rho=\pm (1-2e^{-x})$ 
corresponding to a  parametrization
from the case {\bf T${}_\text{c}$}, we do not reproduce 
the last case {\bf E}, but obtain, instead, 
$L_0=\frac{1}{4}m(x)\dot{x}^2$ with position-dependent
mass $m(x)$ and $x(\chi)$ functions given by
\be\label{m(x)E}
m(x)= \frac{1}{(e^x-1)k'^2 +4k^2(1-e^{-x})^2}\,,\qquad
x(\chi)=-\ln\left(\frac{1}{2}(1+\sn\, \chi)\right)\,,
\ee
and potentials 
\be\label{u1E}
u_1(x)=4k^2(e^{-2x}-e^{-x})-k'^2\,,
\qquad
u_1(x)=\frac{k'^2+4k^2(e^{-x}-e^{-2x})}{4(e^{-2x}-e^{-x})}\,
\ee
for the cases {\bf L${}_\text{E}$} and {\bf D${}_\text{E}$},
respectively. In the limit $k\rightarrow 0$, 
the mass function from (\ref{m(x)E}) transforms into
that for the case {\bf T${}_\text{c}$}, 
but for $k\rightarrow 1$ we obtain $m(x)=\frac{1}{4(1-e^{-x})^2}$,
which does not appear in Table \ref{Table3}, and, particularly, 
does not coincide with $m(x)$ for the {\bf H${}_\text{c}$} case.
There is no contradiction here since in general 
different elliptic functions 
may have the same limit for $k\rightarrow 0$
(or for $k\rightarrow 1$), but different limits 
as $k\rightarrow 1$ ($k\rightarrow 0$). 
For the discussion of this point 
in application  to  finite-gap systems, see 
\cite{CJNP,CJPFin,PAN}.

The first function from (\ref{u1E}) 
has a form of Morse potential.
So, the system (\ref{m(x)E}) with first potential from  (\ref{u1E})
gives us finite-gap elliptic  generalization of  reflectionless 
system with position-dependent mass
$m(x)=\frac{1}{4(1-e^{-x})^2}$ and 
Morse-like potential $u_1(x)=4(e^{-2x}-e^{-x})$.

\vskip0.2cm

\subsection{Special case of elliptic finite-gap systems
and Bernoulli lemniscate}\label{subBernoulli}

In the limits $k\rightarrow 1$ and $k\rightarrow 0$, 
the elliptic finite-gap systems we considered 
transform into hyperbolic and trigonometric systems.
A rather natural question that appears here whether anything interesting 
happens in the middle case, at $k^2=k'^2=\frac{1}{2}$\,.
{}In this   
case we have $k=k'=\frac{1}{\sqrt{2}}$, and so, $
\text{\bf K}(k)=\text{\bf K}(k')\equiv
\text{\bf K}'(k)$, i.e. in this case the magnitudes
of the real,
$2\text{\bf K}( 1/\sqrt{2})$ and 
the hidden
imaginary,  $2i\text{\bf K}'(1/\sqrt{2})$, periods of finite-gap {\bf L}--
 and  {\bf D}--potentials  $-C_n\,\dn^2\,\chi$ and 
$C_n\,\text{dc}^2\,\chi$  coincide.  
This corresponds to  the  lemniscatic 
case of elliptic functions \cite{WW,NIST} 
with a purely imaginary value $k=i$ of the modular 
parameter
for which 
$
\sn\,(z,i)=\frac{1}{\sqrt{2}}\,\sd\,\left(\sqrt{2}\,z,{1}/{\sqrt{2}}\right)$,
$\cn\,(z,i)=\cd\,\left(\sqrt{2}\,z,{1}/{\sqrt{2}}\right)$ and 
$\dn\,(z,i)=\nd\,\left(\sqrt{2}\,z,{1}/{\sqrt{2}}\right)
$, see Appendix B. In the case $k=i$ for the complementary 
modular parameter we have $k'^2=1-k^2=2$.

In lemniscatic case $k=i$,  
the Hamiltonian operator of finite-gap Lam\'e system 
takes the form
$H^{\text{\bf L}}_n=2\left(-\frac{d^2}{d\zeta^2}-
C_n\dn^2\,(\zeta,1/\sqrt{2}\right)$, where
$\zeta\equiv \sqrt{2}\,\chi +\text{\bf K}\left({1}/{\sqrt{2}}\right)$.
This is just the rescaled  Hamiltonian of the displaced in a half-period 
$n$-gap Lam\'e system  with $k=1/\sqrt{2}$.

For the basic {\bf D}--potential with $k=i$ we have 
$\text{dc}^2\,(\chi,i)=\text{nc}^2(\sqrt{2}\,\chi,1/\sqrt{2})=
-2\dn^2\,(\zeta,1/\sqrt{2})+1$, $\zeta\equiv\sqrt{2}\,\chi+
\text{\bf K}(1/\sqrt{2})+i\text{\bf K}'(1/\sqrt{2}))$, and  the Hamiltonian 
is rewritten equivalently $H_n^{\text{\bf D}}=2\left(-\frac{d^2}{d\zeta^2}-
C_n\dn^2\,(\zeta,1/\sqrt{2})\right)+1$.
This is a rescaled finite-gap {\bf D}-Hamiltonian operator. 

Let us show now that all the {\bf L}- and {\bf D}- finite-gap systems
with position-dependent mass presented in Table \ref{Table3},
in the   lemniscatic case $k=i$ can be obtained from 
a non-relativistic particle of mass $m=1$ in Eucledian space $\R^2$
with coordinates $(\xi,\eta)$, which is subjected to the action of  one of 
the two  basic potentials 
\be\label{uLDlem}
u^{\text{\bf L}}_1(\xi,\eta)=-\frac{2}{\xi^2+\eta^2+1}\,,\qquad
u^{\text{\bf D}}_1(\xi,\eta)=\frac{1}{\xi^2+\eta^2}\,
\ee
and restricted to move along  the Bernoulli lemniscate. 

\vskip0.1cm

 \begin{figure}[htbp]
\begin{center}
\includegraphics[scale=0.6]{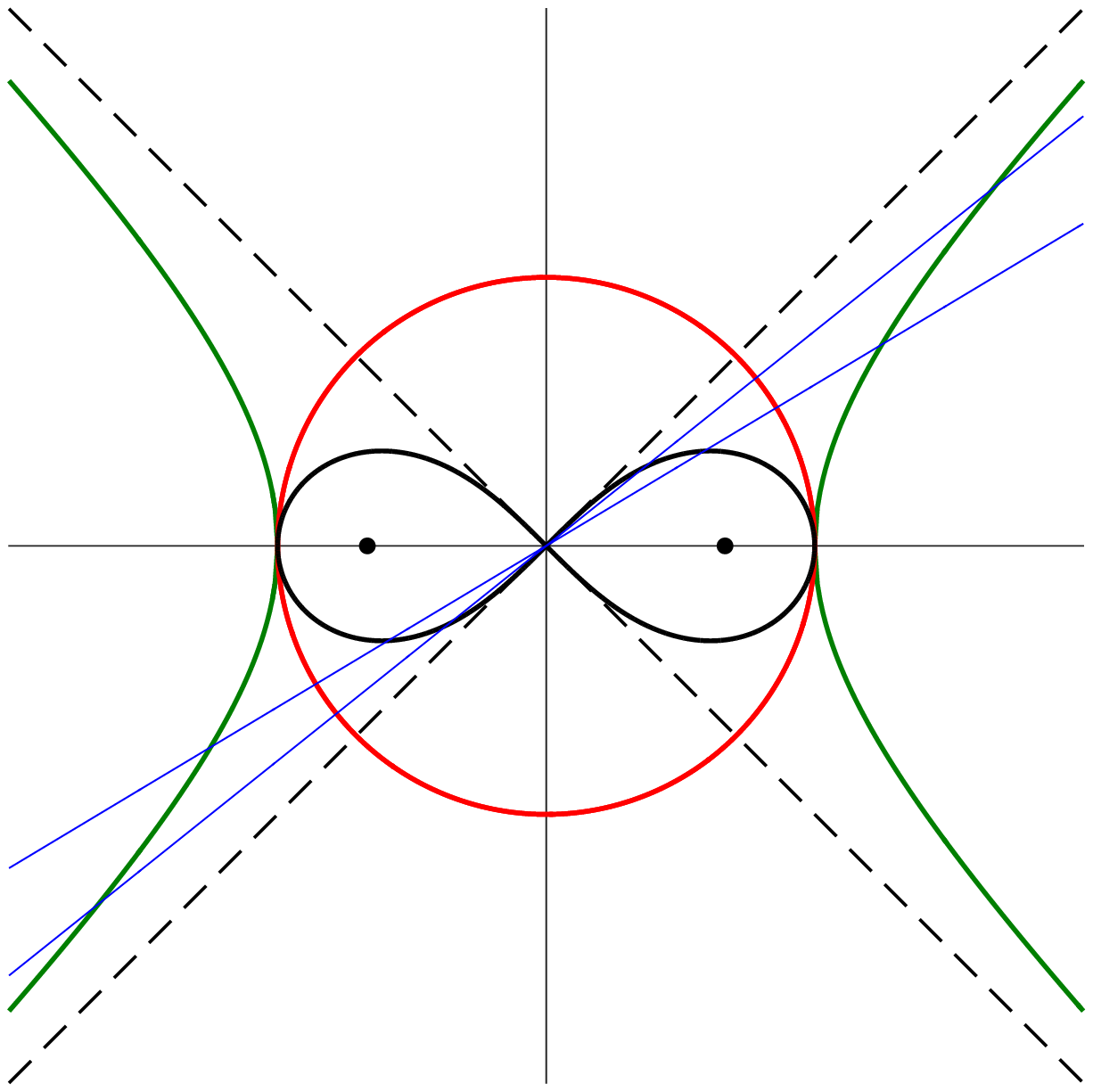}
\caption{Bernoulli lemniscate, equilateral hyperbola and 
circumference of inversion.}
\label{Lem}
\end{center}
\end{figure}

Bernoulli lemniscate can be obtained in the following 
way directly relevant  to our consideration.
Take an  equilateral (rectangular)  hyperbola
in Euclidean $\R^2$ space given by the 
equation  $X^2-Y^2=1$,
and construct its inversion in the circle of unit radius centered
at the origin of the system of coordinates, see Figure \ref{Lem}.
We obtain 
$
(X,Y)\longmapsto (\xi,\eta)\,,
$
where
\be\label{xietaXY}
(\xi,\eta)=\left(\frac{X}{X^2+Y^2}\,,\,\frac{Y}{X^2+Y^2}\right)\,.
\ee
We have $\xi^2+\eta^2=1/(X^2+Y^2)$, $\xi^2-\eta^2=1/(X^2+Y^2)^2$,
and so, the points of the inverted hyperbola satisfy the equation
$$
(\xi^2+\eta^2)^2=\xi^2-\eta^2\,.
$$
This is nothing else as the equation of a particular case of the 
Bernoulli lemniscate 
$(\xi^2+\eta^2)^2=2c^2(\xi^2-\eta^2)$
with foci at $(-c,0)$ $(+c,0)$ with $c=1/\sqrt{2}$,  and unit 
`radius' $\sqrt{2}\,c=1$.

For a particle restricted to move on the lemniscate, we have 
\be\label{paramxi}
\dot{\xi}^2+\dot{\eta}^2=\frac{\dot{Y}^2}{(1+Y^2)(1+2Y^2)}\,,
\ee
where we have taken  into account that $X^2-Y^2=1$. 
Taking now all the different  parametrizations  for the equilateral hyperbola 
presented in Table \ref{Table1} (with the change $X\leftrightarrow Y$), we reproduce all the
finite-gap elliptic systems with $k=i$  presented in Table \ref{Table3}.
The correspondence between the parametrization  cases 
and elliptic (lemniscatic) families is the following:
\be\label{T1T3corr}
 \text{\bf H}{}_\text{\bf 1} \rightarrow \text{\bf D},\qquad
  \text{\bf H}{}_\text{\bf a}\rightarrow \text{\bf B},\qquad
 \text{\bf H}{}_\text{\bf b} \rightarrow \text{\bf A},\qquad
 \text{\bf H}{}_\text{\bf c} \rightarrow \text{\bf E},\qquad
 \text{\bf H{}}_\text{\bf d}\rightarrow \text{\bf C},\qquad
 \text{\bf H{}}_\text{\bf e} \rightarrow \text{\bf 1}\,.
 \ee
For instance, using  the parametrization from the case 
 $\text{\bf H}{}_\text{\bf 1}$, we put $Y=\sinh x$,
 and (\ref{paramxi}) gives us the kinetic term 
 $L_0=\frac{1}{4}m(x)\dot{x}^2$ with $m(x)=1/(1+2\sinh^2 x)$,
 that coincides with the position-dependent mass 
 from  Table \ref{Table3} for the $ \text{\bf D}$ family 
 in the lemniscatic case $k=i$.
 Potentials (\ref{uLDlem}) take here the form 
 of the potentials of the lemniscatic 
 $ \text{\bf D}$ family: 
$u^{\text{\bf L}_{\bf D}}_1=-2+\cosh^2 x$ 
and 
$u^{\text{\bf D}_{\bf D}}_1=1+2\sinh^2 x$.  
Using the parametrization for the case 
 $\text{\bf H}{}_\text{\bf e}$, we put $Y=\frac{\sn\,(x,k)}{\cn\, (x,k)}$,
 and (\ref{paramxi}) gives us $m(x)=\dn^2\,(x,k)/(1+\sn^2\,(x,k))$.
 For lemniscatic case $k=i$ this reduces to the constant mass $m=1$ 
 for the family $\text{\bf 1}$ from  Table  \ref{Table3},
 while (\ref{uLDlem})  gives us the corresponding potentials.
 In a similar  way, one can check other correspondences 
 shown in Eq. (\ref{T1T3corr}).
 Particularly, parametrization $Y=\frac{x}{\sqrt{1-x^2}}$
 from $\text{\bf H}{}_\text{\bf b}$ case gives the mass function
 $m(x)=\frac{1}{1-x^4}$ of the lemniscatic case for  the family
 {\bf A}.  Since for $\text{\bf H}{}_\text{\bf a}$ case we have 
 $Y=x$, Eq. (\ref{paramxi}) gives us immediately the mass function
 for the lemniscatic {\bf B} family: $m(x)=\frac{1}{(1+x^2)(1+2x^2)}$.
The only correspondence that requires an additional step to establish 
is  $\text{\bf H}{}_\text{\bf c} \rightarrow \text{\bf E}$.
 The parametrization from the case  $\text{\bf H}{}_\text{\bf c}$
 with $X=\frac{1}{2}(\kappa+\frac{1}{\kappa})$,  $Y=\frac{1}{2}(\kappa-\frac{1}{\kappa})$
  gives us the rational 
 parametrization (\ref{xietaXY}) of the lemniscate,
 $$
 (\xi,\eta)=\left(\frac{\kappa^3+\kappa}{\kappa^4+1},
 \frac{\kappa^3-\kappa}{\kappa^4+1}\right),
 $$
 and the kinetic term $L_0(\kappa)=\frac{1}{4}m(\kappa)\dot{\kappa}^2$ 
 with the position-dependent mass $m(\kappa)=2/(1+\kappa^4)$.
 Changing additionally the parameter by
 $\kappa=\sqrt{\sinh \tau}$, $\tau\equiv \sqrt{2}\,x$,
 where we assume $x>0$, 
 the obtained kinetic term transforms into 
 $L_0(x)=\frac{1}{4}m(x)\dot{x}^2$ with $m(x)=\frac{1}{\sinh\tau}$,
 that corresponds to the lemniscatic case $k=i$ of the 
 position-dependent mass for the finite-gap 
 family {\bf E} from the Table \ref{Table3}.
 It is not difficult to check also that Eq. (\ref{uLDlem}) 
 reproduces correctly 
 the lemniscatic form of the potentials 
 for the same family  {\bf E}.

\vskip0.3cm

\subsection{Finite-gap systems
  by reduction of a free particle on surfaces of revolution}\label{revolution}

We have showed how hyperbolic, trigonometric and 
elliptic finite-gap systems with PDM can be 
obtained  by appropriate reduction procedures 
in different spaces  of constant curvature
in the presence of Calogero-like or harmonic 
oscillator potentials, or potentials related to these 
ones via appropriate coordinate transformations.
Here we discuss how the same systems can be generated by 
the angular momentum reduction of a free particle system 
on some surfaces of revolution.

Hyperbolic finite-gap systems  can be obtained 
by taking a free non-relativistic particle on one-sheet hyperboloid 
embedded into (2+1)-dimensional Minkowski space $\R^{1,2}$,
 the 
 $AdS_2$ space, 
and by making a phase space reduction 
of the system to a surface of a constant angular momentum.
Indeed, 
 consider a one-sheet hyperboloid  with coordinates 
$x^0=X(x)$, $\vec{x}=Y(x)\vec{n}(\varphi)$,
where $Y(x) =\sqrt{1+X^2(x)}$, $-\infty<X(x)<\infty$, 
$\vec{n}(\varphi)=(\cos\varphi,\sin\varphi)$,
$0\leq \varphi <2\pi$.  We assume the hyperboloid 
is imbedded into the (2+1)-dimensional 
Minkowski space with metric $\eta_{\mu\nu}=\text{diag}\,(+1,-1,-1)$. {}From a free particle  Lagrangian 
$L=\frac{1}{4}\eta_{\mu\nu}\dot{x}^\mu\dot{x}^\nu=
\frac{1}{4}((\dot{x}^0)^2-(\dot{\vec{x}})^2)$
in  $\R^{1,2}$ 
we obtain the Lagrangian 
 $L=\frac{1}{4}(m(x)\dot{x}^2-Y^2(x)\dot{\varphi}^2)$,
where $m(x)=X'^2/(1+X^2)=Y'^2/(Y^2-1)$. 
By the construction, the system is $SO(2,1)$-invariant.
The angular coordinate $\varphi$ is cyclic, 
and  corresponding  Routhian is
$R=\frac{1}{4}\left(m(x)\dot{x}^2-p_\varphi^2\frac{1}{Y^2(x)}\right)$.
Conserved canonical momentum $p_{\varphi}=-\frac{1}{2}Y^2(x)\dot{\varphi}$ 
is the angular momentum of the system
generating rotations in the plane $\vec{x}\in\R^2$, and 
reduction of the system to the surface
 $p_\varphi=C$ 
corresponds to the Lagrangian
$L=\frac{1}{4}\left(m(x)\dot{x}^2-C^2\frac{1}{Y^2}\right)$
considered  by us when we started our discussion 
of the hyperbolic family of the systems.
On the other hand, the reduction can be realized at the 
quantum level in such a way that the 
quantum constant $C_n=n(n+1)\hbar^2$
will be reproduced exactly in the emerging
potential term.
For this it is necessary to introduce into initial 
Lagrangian a topologically
nontrivial term $-\alpha\dot{\varphi}$, which does not change 
classical equations of motion and corresponds to 
coupling of the particle  to the Aharonov-Bohm flux
\cite{CJPAdS}.

In analogous way, one can  obtain  trigonometric finite-gap systems
by considering a free particle on a sphere embedded in 3D Euclidean space,
$x_3=X(x)$, $\vec{x}=Y(x)=\vec{n}(\varphi)$, where 
$X^2+Y^2=1$, $Y>0$, and $\vec{n}(\varphi)$ is the unit vector
as in the hyperbolic case. 
Then $L=\frac{1}{4}(\dot{x}^3)^2+(\dot{\vec{x}})^2=
\frac{1}{4}\left(m(x)\dot{x}^2+Y^2(x)\dot{\varphi}^2\right)$,
$m(x)=\frac{Y'^2}{1-Y^2}$. 
Analogously to the previous hyperbolic case,
by reduction to the surface of the constant angular 
momentum $p_\varphi$ one can  reproduce finite-gap
trigonometric systems.

One can also consider a free motion of the particle 
on upper (or lower) sheet of the two-sheeted hyperboloid 
$(x^0)^2-(\vec{x})^2=1$ embedded into 
the three-dimensional Minkowski 
space~\footnote{
Stereographic projection of one sheet
of a two-sheeted hyperboloid embedded into 
$\R^3$ gives the Poincar\'e disc model of 
Lobachevsky plane, where
the appropriate reduction 
of the kinetic term on geodesics, as we have seen,   
supplies us with the kinetic terms for the systems
of {\bf H} family. 
}
 $\R^{2,1}$.
Taking the upper sheet given by $x^0=X(x)=\sqrt{1+Y^2(x)}$, 
$\vec{x}=Y(x)\vec{n}(\varphi)$
 and a free particle 
Lagrangian in the  form 
 $L=\frac{1}{4}((\dot{\vec{x}})^2-(\dot{x}^0)^2)$, we 
 obtain $L=\frac{1}{4}m(x)\dot{x}^2+Y^2(x)\dot{\varphi}^2$ and 
 $R=\frac{1}{4}m(x)\dot{x}^2+p_{\varphi}^2\frac{1}{Y^2(x)}$,
 $m(x)=\frac{Y'{}^2}{1+Y^2}$.
 After reduction to the surface $p_{\varphi}=C$ of a constant value of the integral of motion
 $p_{\varphi}$, one can reproduce  singular finite-gap 
 hyperbolic systems. Particularly, the choice $Y(x)=\frac{1}{\sqrt{x^2-1}}$, $x>1$,
 reproduces the system {\bf  H${{}{\bf '}}_\text{\bf b}$} 
 with $m(x)=\frac{1}{(x^2-1)^2}$ and $u_1(x)=x^2-1$, which 
 after the change of variable $x\rightarrow \chi$, $x=-\coth \chi$, $\chi<0$,
 transforms this into the system  {\bf  H${{}{\bf '}}_\text{\bf 1}$} with $m=1$ and $U_1(\chi)=\frac{1}{\sinh^2 \chi}$. 
 The choice $Y(x)=\sqrt{x^2-1}$, $x>1$, reproduces the system
 {\bf  T${{}{\bf '}}_\text{\bf a}$} with $m(x)=\frac{1}{x^2-1}=u_1(x)$,
which after the change of variable 
gives $m(\chi)=1$ and   $U_1(\chi)=\frac{1}{\sinh^2 \chi}$.

%%%%%%%%%%%%%%%%%%%
  The both hyperbolic reflectionless  {\bf H} and finite-gap singular 
{\bf H}${}'$
families can be obtained from ordinary 
Lorentzian anti-De Sitter spacetime $AdS_3$ of curvature  radius $\ell$
by treating it as being embedded in $\R^{2,2}$.
The embedding is given by the equation
$-(\vec{x}_-)^2+(\vec{x}_+)^2=-\ell^2$, where
$\vec{x}_\pm$ are two-dimensional vectors
with components which we denote by $x^{1,2}_\pm$.
Parametrization 
$\vec{x}_-=\ell \cosh \rho\, \vec{n}_-(\tau/\ell)$,
$\vec{x}_+=\ell \sinh \rho\, \vec{n}_+(\varphi)$,
$\vec{n}_\pm=(\cos\lambda_\pm,\sin\lambda_\pm)$,
$\varphi\in[0,2\pi)$, $\tau\in[0,2\pi\ell)$, $\rho\in[0,\infty)$
gives the $AdS_3$ metric 
\be
ds^2=-\cosh^2\rho\, d\tau^2 +\ell^2\sinh^2\rho \, d\varphi^2
+\ell^2 d\rho^2\,.
\ee
Taking a free particle in $AdS_3$ described by  Lagrangian 
$L_0=\frac{1}{4}(ds/dt)^2=\frac{1}{4}(-\cosh^2\rho \,\dot{\tau}^2 +
\ell^2\sinh^2\rho \, \dot{\varphi}^2
+\ell^2 \dot{\rho}^2)$, 
the Hamiltonian reduction by constraints
 $p_\varphi=C_1$, $p_\tau=0$ provides us with
 singular finite-gap systems {\bf H}${}_{\bf 1}'$.
 Instead of the second condition (constraint) one can 
 take $\tau=0$, that corresponds to restriction on the subspace
 $x^2_-=0$, $x^1_-\geq \ell$.
To obtain the reflectionless {\bf H}${}_{\bf 1}$-family,
we reduce the system 
by using the constraints $p_\tau=C$, 
$x^2_+=0$. 
In the subspace with $x^2_+=0$ we have $x^1_+\geq 0$
that corresponds to $\varphi=0$, and 
$x^1_+\leq  0$
for $\varphi=\pi$.
These two subspaces can be unified 
by taking $\varphi=0$ and extending 
$\rho$ from $[0,\infty)$ to 
the infinite interval $(-\infty,\infty)$.
Such extension (doubling) of the interval
for the variable $\rho$ is similar to the picture 
taking place for the motion along the Seiffert spiral
we utilized  to generate finite-gap elliptic systems. 
Again, by appropriate change of the variable $\rho$,
we reproduce all the hyperbolic finite-gap systems
with position-dependent mass we discussed.
\vskip0.1cm

%%%%%%%%%%%%%%%%

One can also obtain  elliptic systems {\bf L} by taking a free 
particle on a certain surface of revolution embedded 
into Minkowski (2+1)-dimensional space. 
For the family $\text{\bf L}{}_\text{\bf 1}$,
the corresponding surface in a two-parametric form is 
given by $x^0=\frac{1}{k'}\text{E}(\xi,k)$,
$\vec{x}=\frac{1}{\dn\,\chi}\vec{n}(\varphi)$,
where $\xi=k'\text{sc}\,(\chi,k)=-\cn\,(\chi+\text{\bf K},k)$,
and $\text{E}(\xi,k)$ is the incomplete elliptic integral of the 
second kind, $\text{E}(x,k)=\int_0^x\sqrt{\frac{1-k^2\tau^2}{1-\tau^2}}d\tau$.
This surface represents a surface of a form of a one-sheet hyperboloid 
but with $-\text{\bf E}\leq x^0\leq \text{\bf E}$, where 
$ \text{\bf E}=\text{E}(1,k)$ is the complete elliptic integral of the 
second kind \cite{WW}.
In the limit $k\rightarrow 1$, this surface transforms into 
the one-sheeted hyperboloid ($AdS_2$)
surface we discussed above, while in the another  limit $k\rightarrow 0$,
it transforms into a cylinder  with $-\pi\leq x^0\leq \pi$.
We have here  $L_0=\frac{1}{4}((\dot{x}^0)^2-(\dot{\vec{x}})^2)=
\frac{1}{4}(\dot{\chi}^2-\nd^2\,\chi\,\dot{\varphi}^2)$.
After reduction to the surface $p_\varphi=C$, 
this yields the $\text{\bf L}{}_\text{\bf 1}$ family of the systems
with $m=1$.
Other {\bf L}--families
of the  systems  
with position-dependent mass we discussed 
can be obtained by the change of variable using the information
presented in Table \ref{Table3}.
By a complex displacement $\chi \rightarrow 
\chi+\text{\bf K}+i\text{\bf K}'$, one can also generate
the {\bf D}--families of finite-gap systems.

\section{Supersymmetric pairs of  
finite-gap
systems}\label{SecSUSY}

Consider now some examples of supersymmetric finite-gap systems
with position-dependent mass which can be obtained based 
on the constructions of the preceding  sections. 

Let us take a function $\Phi(\chi)$ to be nodeless in a certain 
interval $(\chi_1,\chi_2)$. 
In accordance with  (\ref{superpot}), 
in a usual way we obtain a  superporpotential $\mathcal{W}(\chi)=\Phi'(\chi)/\Phi(\chi)$
to be non-singular function in the same interval.
In terms of $\mathcal{W}(\chi)$  we construct two quantum systems,
$H_{\Phi}$, defined by Eq. (\ref{HPhi}), and $H_{1/\Phi}$.
They form a supersymmetric pair  
$(H_+\equiv H_{1/\Phi},\,  H_-\equiv H_{\Phi})$, 
$H_\pm=-\hbar^2 \frac{d^2}{d\chi^2}+\mathcal{V}_\pm$, with potentials 
$\mathcal{V}_\pm=\mathcal{W}^2\pm \hbar \mathcal{W}'$. 

The choice 
\be\label{PhicoshH}
\Phi=(\cosh \chi)^n\,,\qquad
\mathcal{W}=\hbar n\tanh\chi\,,\qquad \chi\in(-\infty,\infty)\,,
\ee
with $n=1,2,\ldots$, gives us a supersymmetric pair of quantum systems 
with 
\be\label{VpmH}
\mathcal{V}_\pm=\hbar^2n^2-C_{\mp n}\,\frac{1}{\cosh^2\chi}\,,\qquad
C_{\pm n}=\hbar^2n(n\pm1)\,,
\ee
where $C_{-n}=C_{n-1}$. 
We show explicitly  the dependence on Planck constant
to stress the purely quantum nature of the  potentials.  
These are the pairs of  reflectionless hyperbolic 
systems with $n$ bound states in the system $H_-$
and $n-1$ bound states in $H_+$, 
where $H_+$ at $n=1$ corresponds to 
a free particle on a real line.

Analogously, we obtain the supersymmetric pairs 
of finite-gap trigonometric systems,
\be\label{PhicosT}
\Phi=(\cos \chi)^n\,,\qquad
\mathcal{W}=-\hbar n \tan\chi\,,\qquad
\chi\in(-{\pi}/{2},{\pi}/{2})\,,
\ee
\be\label{VpmT}
\mathcal{V}_\pm=-\hbar^2n^2+C_{\mp n}\,\frac{1}{\cos^2\chi}\,,
\ee
with the basic function $\Phi$ to be nodeless
 in the indicated finite interval.

The choice 
\be\label{PhicosE}
\Phi=(\dn\,\chi)^n\,,\qquad
\mathcal{W}=-\hbar n k^2\,\frac{\sn\,\chi\cn\,\chi}{\dn\,\chi}\,,\qquad
\chi\in(-\infty,\infty)\,
\ee
with the periodic basic function $\Phi$ to be nodeless on all the real line
produces the supersymmetric pair of the systems with potentials 
\be\label{VpmE}
\mathcal{V}_\pm=\hbar^2 n^2(1+k'^2) -C_{\mp n}\dn^2\chi
-C_{\pm n}\dn^2\,(\chi+\text{\bf K})\,,
\ee
where $\dn\,(\chi+\text{\bf K})={k'}/{\dn\,\chi}$. At $n=1$, 
(\ref{VpmE}) corresponds to a pair of one-gap Lam\'e systems
with potentials mutually shifted in the half 
of their real period. For $n>1$, these are the supersymmetric pairs of
$n$-gap associated Lam\'e systems of a special form \cite{CJNP,CJPFin}, see below.
In the infinite period limit corresponding to $k\rightarrow 1$, they transform into the 
supersymmetric hyperbolic pairs (\ref{VpmH}), while for $k\rightarrow 0$
both potentials turn into zero. The supersymmetric partner
potentials (\ref{VpmE}) satisfy the property
\be\label{persift}
\mathcal{V}_\pm(\chi+\text{\bf K})=\mathcal{V}_\mp(\chi)\,,
\ee
which means that the corresponding supersymmetric partner
Hamiltonians $H_+$ and $H_-$ are completely isospectral. 
By the construction, the functions 
$\Phi(\chi)=(\dn\,\chi)^n$ and $1/\Phi(\chi)$ are the eigenstates 
of the $H_+$ and $H_-$ systems, respectively.
They  correspond to non-degenerate ground states 
of zero energy of these systems \cite{CJNP,CJPFin}.

By the complex shift 
$\chi\rightarrow \chi+i\text{\bf K}'$ in  (\ref{PhicosE}) and (\ref{VpmE}) 
we obtain the analog
which describes singular supersymmetric systems 
belonging to   the family of Darboux-Treibich-Verdier   finite-gap systems:
\be\label{PhicosEcs}
\Phi=(\text{cs}\,\chi)^n\,,\qquad
\mathcal{W}=-\hbar n \,\frac{\dn\,\chi}{\sn\,\chi\,\cn\,\chi}\,,\qquad
\chi\in(0,\text{\bf K})\,,
\ee
\be\label{VpmEcs}
\mathcal{V}_\pm=\hbar^2 n^2(1+k'^2) +C_{\mp n}\,\text{cs}^2\,\chi
+C_{\pm n}\,k'^2 \text{sc}^2\,\chi\,.
\ee
The last  term in (\ref{VpmEcs}) can be presented equivalently 
 in the form $C_{\pm n}\, \text{cs}^2\,(\chi+\text{\bf K})$, that can be compared 
 with the structure in (\ref{VpmE}). As a consequence,
the  superpartner potentials  in (\ref{VpmEcs})  satisfy the property (\ref{persift}).
In the limit $k\rightarrow 1$, (\ref{VpmEcs}) transforms into
supersymmetric pair of singular hyperbolic systems described by potentials
$\mathcal{V}_\pm=\hbar^2n^2+C_{\mp n}\frac{1}{\sinh^2\chi}$,
while in another limit $k\rightarrow 0$ we obtain supersymmetric pairs
with partner potentials $\mathcal{V}\pm=2\hbar^2n^2+C_{\mp n}\,\text{cotan}^2\,\chi
+C_{\pm n}\,\text{tan}^2\,\chi$  
\cite{SUSYQM}.

To construct finite-gap elliptic supersymmetric system
which in trigonometric limit $k\rightarrow 0$
reproduces  supersymmetric finite-gap  family 
(\ref{PhicosT}), (\ref{VpmT}), we make in (\ref{PhicosE}), (\ref{VpmE}) a change
$\chi\rightarrow i\chi$, multiply the resulting Hamiltonian operators 
by $-1$, and make  a change 
$k\leftrightarrow k'$.  
This yields 
\be\label{PhicosEdc}
\Phi=(\text{dc}\,\chi)^n\,,\qquad
\mathcal{W}=-\hbar n k'^2\,\frac{\sn\,\chi}{\cn\,\chi\,\dn\,\chi}\,,\qquad
\chi\in(-\text{\bf K},\text{\bf K})\,,
\ee
\be\label{VpmEdc}
\mathcal{V}_\pm=-\hbar^2 n^2(1+k^2) +C_{\mp n}\,\text{dc}^2\,\chi
+C_{\pm n}k^2\,\text{cd}^2\,\chi\,.
\ee
Potentials (\ref{VpmEdc}) satisfy, again,  the property 
(\ref{persift}). The limit $k\rightarrow 1$ applied to  
(\ref{VpmEdc}) gives  
$\mathcal{V}_\pm=0$, while in another limit $k\rightarrow 0$
we reproduce supersymmetric trigonometric pair (\ref{VpmT}). 
Note that the last term in (\ref{VpmEdc}) can be written equivalently as
$C_{\pm n}\text{dc}^2\,(\chi+i\text{\bf K}')$, that can be compared 
with the properties  of separate terms of superpartner potentials 
in (\ref{VpmE}) and (\ref{VpmEcs}) under 
the real displacement $\text{\bf K}$. In correspondence 
with this, the potentials in (\ref{VpmEcs})  can be presented 
equivalently  
$\mathcal{V}_+(\chi)=-\hbar^2 n^2 k^2 +C_{n}\,\text{dc}^2\,\chi
+C_{-n}\, \text{dc}^2\,(\chi+\text{\bf K})=V_-(\chi+\text{\bf K})$. 
This 
 particularly explains  the following seeming paradox.
As we saw in the previous section,
in the non-supersymmetric case 
 the finite-gap Lam\'e and singular elliptic systems, 
which in the limits
$k\rightarrow 1$ and $k\rightarrow 0$  produce finite-gap
hyperbolic and trigonometric systems, can be 
related either via the complex shift $\chi\rightarrow \chi+\text{\bf K}
+\text{\bf K}'$ or via the the transformation 
$\chi\rightarrow i\chi$. 
However,
 these two types of transformations 
applied to supersymmetric
associated Lam\'e system (\ref{VpmE}) produce two different supersymmetric 
systems belonging to the Darboux-Treibich-Verdier 
families of finite-gap systems.

In all the supersymmetric families of finite-gap systems presented above, mass is 
position-independent, $m=1$. To reconstruct the 
supersymmetric systems 
with position-dependent mass, consider 
as a first  example   the mass function $m_{{}_{\bf A}}(x)=1/(1-x^2)(1-k^2x^2)$ 
corresponding to the elliptic family {\bf A} from the previous section.
{}From Table \ref{Table3} 
we find that in this case  the function (\ref{varphifx}) giving  the change of 
variable is $\varphi(\chi)= \cn\,\chi\dn\,\chi$, and $x(\chi)=\sn\,\chi$.
Eq. (\ref{Phisxigen}) allows us  to find the  functions
$\varsigma(x)$ for supersymmetric pair of finite-gap systems given by 
potentials (\ref{VpmE}). We denote these functions by
$\varsigma_\pm(x)$, and obtain  $\varsigma_\pm(x)=(1-k^2x^2)^{\mp \frac{n}{2}-
\frac{1}{4}}(1-x^2)^{-\frac{1}{4}}$, 
$x\in(-1,1)$. 
The supersymmetric pair of $n$-gap
quantum systems 
$H_+(x)$ and $H_-(x)$  with position-dependent 
mass $m_{{}_{\bf A}}(x)$
 is reconstructed then with the help of Eq.  (\ref{Hetaf}), 
where  $f(x)=1/\sqrt{m_{{}_{\bf A}}(x)}$.
In the limit $k\rightarrow 1$, we have
$f(x)=1-x^2$, $\varsigma_\pm=(1-x^2)^{-\frac{1}{2}(1\pm n)}$,
and obtain supersymmetric pair of reflectionless
systems of the type considered by Linde et al
\cite{Linde}.

Consider now another example of
 the position-dependent mass function $m_{{}_{\bf C}}=1/(1-k^2\sin^2 x)$,
$x\in(-\infty,\infty)$,  corresponding
to the family {\bf C} in Table \ref{Table3}. 
In this case  the change of variable function 
$\varphi(\chi)=\dn\,\chi$ is related to the supersymmetry-generating function
$\Phi$ from (\ref{PhicosE}) in a simple exponential  way. This allows
us to  use the ordering prescription corresponding to
the similarity transform  function (\ref{sigflam}) given in terms 
of the mass function. The  parameter $\lambda$ 
for corresponding superpartner potentials (\ref{VpmE})  
is fixed in the form $\lambda_\pm=\frac{1}{2}\pm n$, and
here, as follows from Table \ref{Table3}, $\sn\,\chi=\sin x$,
and $f(x)=1/\sqrt{m_{\bf C}(x)}=\sqrt{1-k^2\sin^2 x}$.
In 
accordance 
with (\ref{Hlam-lam}), 
the supersymmetric pair of finite-gap 
systems corresponding to the pair of associated Lam\'e 
systems (\ref{PhicosE})  is given by Hamiltonian operators
with position dependent mass,
\be
H_\pm(x)=-\hbar^2 m_{{}_{\bf C}}^{\mp \frac{n}{2}-\frac{3}{4}} \frac{d}{dx}
m_{{}_{\bf C}}^{\pm n+\frac{1}{2}}\frac{d}{dx}m_{{}_{\bf C}}^{\mp \frac{n}{2}-\frac{3}{4}}\,.
\ee
In the limit $k\rightarrow 1$, this pair transforms into a supersymmetric 
pair of reflectionless systems of the type {\bf H}${}_{\bf d}$ presented 
in Table \ref{Table2}.
Since the change of variable function $\varphi(\chi)= \cn\,\chi\dn\,\chi$
from the family  {\bf A} we discussed in the previous example 
and the supersymmetry-generating function $\Phi$ from the
family of the systems  (\ref{PhicosEdc}), (\ref{VpmEdc})
are related as $\Phi=(\varphi(\chi))^{-n}$, one can apply 
the same ordering scheme with generating function (\ref{sigflam})
in this case as well to reproduce kinetic term with position-dependent 
mass which generates supersymmetric finite-gap pairs of the systems
(\ref{VpmEdc}).

Let us stress that in the way described above we 
generate supersymmetric pairs of finite-gap systems
from the kinetic term with position-dependent mass, not introducing 
apart any  potential term. In this sense the construction 
is somewhat reminiscent of the picture of generation
of finite-gap systems via angular momentum reduction 
of a free motion on the surfaces of revolution 
that  we  discussed 
in Section \ref{revolution}.
But this provokes the question if 
the potential terms can be introduced separately in such a way
that we still have supersymmetric pairs of finite-gap systems. 
This can easily be achieved  by
exploiting the not utilized yet  ordering prescription corresponding
to Eq. (\ref{Hsigalp}) in order to construct a pair of finite-gap systems
related by usual supersymmetry generated by supercharges which 
are  first  order differential operators.
Similarly to (\ref{Hsigalp}), we take
\be\label{Halp}
H^U_{+\alpha} \equiv  \frac{1}{2}(1+\alpha)H_{1/\Phi}+\frac{1}{2}(1-\alpha)H_\Phi
+U(\chi)
=
-\hbar^2\frac{d^2}{d\chi^2}+\mathcal{W}^2+\alpha \hbar\mathcal{W}'
+U(\chi)\,,
\ee
with some still unknown potential $U(\chi)$,
and demand that the pair 
$H^U_{+\alpha}$ and $H^U_{-\alpha}$ 
would be supersymmetric. This means  
that the Hamiltonian operators
have to be representable in the form
$H^U_{\pm\alpha}=-\hbar^2\frac{d^2}{d\chi^2}+\mathcal{W}_\alpha^2\pm \alpha \hbar
\mathcal{W}'_{\alpha}$
with some superpotential $\mathcal{W}_\alpha$. 
Equating this with (\ref{Halp}) and
 its analog with $\alpha$ changed 
for $-\alpha$, we find that $\mathcal{W}_\alpha$
can be taken in the form $\mathcal{W}_\alpha=\alpha \mathcal{W}$, where 
for simplicity we set integration constant equal to zero, 
and then $U(\chi)=(\alpha^2-1)\mathcal{W}^2$.
This can be transferred  to 
the case with position-dependent kinetic term 
using the procedure described above.  

What we obtained based on  (\ref{Halp}) is, however, a rather trivial generalization.
We develop it further by considering concrete examples
to generalize for the case of nonlinear supersymmetries 
based on existence of intertwining operators
which are  differential operators of higher order.
Though such a generalization can be realized 
on the basis of the  ordering presented 
in (\ref{Halp}),
we return to the ordering 
we discussed before (which corresponds  to $\alpha^2=1$).  
Let us consider first the concrete 
example of the `inflationary model'  {\bf H}${}_{\bf b}$
with $m(x)=\frac{1}{(1-x^2)^2}$,
 $f(x)=1-x^2$,
$\varphi(\chi)=\frac{1}{\cosh^2\chi}$,
$x(\chi)=\tanh\chi$, and choose the ordering prescription 
based on $\varsigma=f^{-\lambda}$ corresponding to 
(\ref{sigflam}). 
Take the pair
of the quantum systems with quantum kinetic terms 
of the form (\ref{Hlam-lam}), and supply them 
by the potential term of the form $u(x)=\gamma x^2$.
We obtain two-parametric  systems
\be\label{Hpair-lam}
	H_{\lambda,\gamma}(x)\equiv -f^{1-\lambda}\frac{d}{dx}
	f^{2\lambda}\frac{d}{dx}f^{1-\lambda}+\gamma x^2\,,\quad
	\text{and} \qquad H_{1-\lambda,\gamma}
\ee
as two different 
quantum analogs of zero-dimensional
version of the classical field system (\ref{Linde}).
According to (\ref{Phix1/2}), we have $\Phi_{(\lambda)}=
(\cosh\chi)^{2\lambda-1}$ and 
$W=(2\lambda-1)\tanh\chi$. We denote $\beta=2\lambda-1$. 
After the similarity transformation and change of variable 
the Hamiltonian operators 
$H_{\lambda,\gamma}(x)$ and $H_{1-\lambda,\gamma}$
transform into the pair
$H_{\pm \beta,\gamma}(\chi)=-\frac{d^2}{d\chi^2}+V_\pm(\chi)
$
with $V_\pm(\chi)=(\beta^2+\gamma)-\frac{1}{\cosh^2\chi}(\gamma+\beta^2\mp
\beta)$, where  we set  again $\hbar=1$. 
Both obtained systems in the pair are reflectionless hyperbolic  systems if
coefficients are chosen such that
$(\gamma+\beta^2\mp \beta)=n_\pm(n_\pm+1)\equiv C_{n_{\pm}}$, where 
$n_+$ and $n_-$ are some integer numbers (with zero value corresponding
to a  free particle  case).
This gives 
$\lambda=\frac{1}{4}(C_{n_-}-C_{n_+}+2)$,
$\gamma=\frac{1}{2}(C_{n_-}+C_{n_+})-\beta^2$,
and  then
$V_\pm(\chi)=\frac{1}{2}(C_{n_-}+C_{n_+}) 
-C_{n_\pm} \frac{1}{\cosh^2\chi}$.
In particular, when one of the integers $n_-$ or $n_+$ is equal to zero,
one of the systems in the pair 
corresponds to the free particle. 
Reflectionless system with 
coefficient $C_{n}=n(n+1)$ in potenial term 
can be related to the free particle  Hamiltonian
by means of intertwining operator which is a
 differential operator 
of order $n$.  Assuming that $n_+>n_-$,
 and since the free particle 
is characterized by the momentum 
operator integral $-i\frac{d}{d\chi}$,  
the  systems with coupling constants 
$C_{n_+}$  and $C_{n_-}$  can be intertwined 
by differential operators of orders
$(n_+ - n_-)$ and $(n_+ + n_- +1)$, 
and the composed system ($H_{n_+}, H_{n_-}$)
will be described by exotic supersymmetry generated
by supercharges of the indicated differential orders and 
by the bosonic integrals composed from 
Lax-Novikov operators of these finite-gap systems,
see \cite{CJNP,CJPFin,AGP} for the details.

In the same way, one can take the pair
(\ref{Hpair-lam}) with  
position-dependent mass $m_{{}_{\bf C}}(x)=\frac{1}{1-k^2\sin^2x}$ corresponding to 
elliptic case we discussed above, and
change the  potential term in (\ref{Hpair-lam}) for 
$u(x)=\gamma(k^2\sin^2x-1)$. 
Then after corresponding similarity transformation  
and the change of variable,  with both operations 
given in terms of 
$f(x)=1/\sqrt{m_{{}_{\bf C}}(x)}$, we find that 
the choice of the parameters 
$\lambda=\frac{1}{2}-l$ and $\gamma=C_n-C_{l-1}$, 
where $l$ and $n$ are integers,
gives us a completely isospectral pair
of the associated Lam\' e systems 
with potentials $V_+(\chi)=-C_n\,\dn^2\,\chi-
C_l\,\dn^2\,(\chi+\text{\bf K})+l^2(1+k'^2)$ and  $V_-(\chi)=V_+(\chi+\text{\bf K})$.
Exotic nonlinear supersymmetry of the system composed from 
Hamiltonians with these associated Lam\'e potentials of the most general form 
is analysed in detail in \cite{CJNP,CJPFin}.

\section{Concluding remarks and outlook}\label{LastSec}

In conclusion, we present below some remarks on the obtained results 
and discuss some interesting problems for future research.

A canonical transformation in the phase space $(x,p)$  generated by a 
function $\alpha(x)$ 
is given by 
\be
g(x,p)\rightarrow {\mathcal G}(x,p)=
\exp \alpha(x)\star g(x,p)\equiv
g(x,p)+\sum_{n=1}^\infty \frac{1}{n!}\{\alpha(x),\{\ldots ,\{\alpha(x),g(x,p)\}\}\ldots
\}\,.
\ee	
Taking a pure imaginary generating function 
 $\alpha(x)=-i\int^x W(\xi)d\xi$, for $g=p$ this yields a complex transformation
$p\rightarrow \mathcal{P}=p-iW(x)$ having a form of a minimal  coupling 
with a purely complex `gauge field' $A(x)=iW(x)$.
In order a transformed kinetic term be real, we take it (in the case $m=1$)  in the form
$h=\bar{\mathcal{P}}\mathcal{P}$, where the bar denotes a  complex conjugation.
Then Hamiltonian operator (\ref{Heta}), (\ref{Heta+})  can also be understood 
as a direct quantum analog of the classical term $h=\bar{\mathcal{P}}\mathcal{P}$.
This picture with a purely complex U(1) `gauge field' $A(x)=iW(x)$ is similar 
to the picture that appears in quasi-exactly solvable systems 
\cite{Shifman}.
{}It seems therefore to be interesting   to  look in more detail for 
relations  between the quantum quasi-exactly solvable systems
and the systems with position-dependent mass.  
Such relations could particularly be relevant in the case of finite-gap systems 
bearing in mind that a hidden 
$so(2,1)$ symmetry plays an important role in understanding 
of their properties \cite{CJNP,CJPAdS}, and that quasi-exact solvability for a broad 
class  of the systems with such a property is based on finite-dimensional
representations  of $sl(2,\R)$ 
\cite{Shifman,Turb,Finkel}. The $sl(2,\R)$ plays also
 important  role 
in the theory of periodic quantum systems 
\cite{CJPFin,integr1}. 

The  kinetic term in (\ref{Heta}), (\ref{Heta+}) 
and then in (\ref{Hetaf}) has a structure similar to that
appearing in  the quantum problem of a particle in 
curved space described by external 
metric  $g_{\mu\nu}(x)$.  
Removal of ordering ambiguity in the quantum kinetic term 
requires there the essential ingredient of invariance 
under general coordinate transformations. 
The same ambiguity   happens 
in flat backgrounds in curvilinear coordinates.  
The  invariance under general coordinate transformations 
is maintained by constructing a quantum kinetic term 
in accordance with the prescription: 
$H=p^-_\mu g^{\mu\nu}p^+_\mu$, where 
$p^\pm_\mu=g^{\pm 1/4}p_\mu g^{\mp 1/4}$ \cite{deWitt}. 
Analogous problem with ordering ambiguity in the kinetic terms 
appears also in the context of supersymmetry 
\cite{deAlf}. 
Let us stress, however, that in both indicated 
cases the analogy  with the present approach to
the quantum mechanical systems with PDM
is rather formal since we considered a one-dimensional case 
here, which is characterized by a trivial 
metric. Nevertheless, 
the fictitious classical similarity transformation in the kinetic term 
we introduced is reminiscent to a freedom 
of the choice of curvilinear coordinates in higher-dimensional flat 
backgrounds. 

We showed that the kinetic term with a position 
dependent mass is a natural source 
to produce the pairs of quantum systems
related by the first order supercharges.
On the other hand, inclusion of the potential term allows us to obtain the pairs 
described by a  nonlinear supersymmetry
with supercharges of arbitrary higher order. 
The  appearance of nonlinear supersymmetry in 
the systems with position-dependent mass deserves 
a further investigation, bearing particularly in mind
a close relation between nonlinear supersymmetry and 
quasi-exact 
solvability  
\cite{CJNP,CJPFin,KliPlyu}.

Though finite-gap systems are described by potentials quadratic in
Planck constant $\hbar^2$, this  does not mean that 
all the systems originating from the 
kinetic term with position-dependent mass as in (\ref{pothbar})
are of this special nature.
On the one hand, finite-gap systems form a very special subclass of the systems
of the form (\ref{pothbar}): they are characterized by the 
presence of a nontrivial Lax-Novikov integral of motion
to be higher order differential operator. The latter,
however,  can be a rather formal integral  in some  quantum 
systems \cite{COP}
unlike the case of integrable systems where it plays a fundamental role 
\cite{integr1,integr2,APChir}.
On the other hand,  nonlinear Riccati  equation 
$\mathcal{W}^2-\mathcal{W}'=V(x)$ with unknown function 
 $\mathcal{W}(x)$ 
always has solutions for arbitrary  
given function  $V(x)$.

A peculiarity of the quantum Bohm potential
$Q$ in the quantum Hamiltonian-Jacobi
equation is that it  is proportional to $\hbar^2$: 
$Q=-\frac{\hbar^2}{2M}\frac{R''}{R}=\frac{\hbar^2}{4M}(\mathcal{S}g)(x)$,
where $R(x)=\sqrt{\rho(x)}$, $g'=\frac{1}{\rho}$, $(\mathcal{S}g)(x)=\frac{g'''}{g'}-\frac{3}{2}\left(
\frac{f''}{f'}\right)^2$ is the Schwarzian, and $\rho(x)$ is the probability density of a quantum state
\cite{QuanPot}.
In (\ref{pothbar}) the potential term 
is $\frac{\hbar^2}{2M}(\mathcal{W}^2-\mathcal{W}')$ that 
coincides with $-Q$ if we make an
identification $\Phi=\frac{1}{\sqrt{\rho}}$. 
In supersymmetric pair (\ref{Hlam-lam}) 
in the case of the ordering prescription with 
$\varsigma=1$  and $\lambda=0$ this identification corresponds to
$f=1/\rho$, while
 for $\varsigma=1$  and $\lambda=1$ one has $f=\rho$.
It would  be interesting to investigate this analogy with the quantum 
Bohm potential in more detail. 
Note that the analogy with the quantum Bohm potential
and its relation to the Schwarzian derivative 
has allowed to one of us to apply the approach 
with  the classically fictitious similarity transformation 
in the kinetic term developed here 
to solve in \cite{SchwarzAn}  
the quantum anomaly problem 
for  supersymmetry with the second-order 
supercharges \cite{KliPlyu}.

The systems with position-dependent mass were studied 
also in the case of spatial dimension $D>1$ \cite{D0,D1,D2,D3,D3+,D4},
particularly, in the context of  superintegrable systems \cite{Nikitin}.
It would be interesting to generalize  
our approach in this direction, having in mind, particularly,
a generalization of Mathiew-Lakshamann model for 
$D>1$ which was studied in  \cite{Carin}.
The analogy with the quantum problem 
of a particle in curved space we noted  above
could  be of important relevance for such 
a generalization. 

Another interesting generalization of the approach presented here 
would be its application to the study of the PT symmetric 
quantum systems.
Some investigations of the systems with PDM 
in the context of
PT symmetry  were realized in
 \cite{PT0,PT1,PT2}.

We showed that some finite-gap periodic elliptic systems
belonging to the broad family of Lam\'e-Darboux-Treibich-Verdier  
systems can be obtained by  reduction to the 
Seiffert's spherical spiral and  Bernoulli's lemniscate
(for a special value of the modular parameter),
or by  angular momentum 
reduction of a free particle motion 
on certain surfaces of revolution 
related to  the  $AdS_2$. 
These observations deserve  a further,
more detailed investigation since in this way
one could expect to obtain  some alternative 
explanation for the origin of  Lax-Novikov integrals 
in finite-gap elliptic  systems by analogy as it was
done for some reflectionless systems 
by considering
 Aharonov-Bohm effect  on $AdS_2$ 
\cite{CJPAdS}.

\vskip0.2cm
\noindent \textbf{Acknowledgements.}
The work has been partially supported by FONDECYT Grant
No. 1130017. 
We  thank  Profs.   J. Cari\~nena and  J. Mateos Guilarte 
for useful discussions.

\section{Appendix A}\label{AppendixA}

Here we show that the quantum kinetic term of the form 
$H_{\alpha,\beta,\gamma}=\mu^\alpha D \mu^\beta D \mu^\gamma +
\mu^\gamma D \mu^\beta D \mu^\alpha$, with $D=\frac{d}{dx}$, 
$\alpha+\beta+\gamma=-1$, $\mu(x)=4m(x)$,
is included into (\ref{Hetaf}) as a particular case.

Equating $H_{\alpha,\beta,\gamma}$ with (\ref{Hetaf}),
we obtain  three relations between coefficients
appearing at $D^2$, $D$ and $D^0=1$. The equality of coefficients at $D^2$ yields 
$f=\sqrt{\frac{2}{\mu}}$. Then the condition  $-2\frac{\mu'}{\mu^2}=2ff'$ which appears as the 
equality of coefficients
at $D$ is satisfied identically. 
Finally, the equality of coefficients at $1=D^0$ can be reduced 
to the equation 
\be\label{Ricc}
(\varsigma'/\varsigma)'-(\varsigma'/\varsigma)^2=\frac{1}{2}(\alpha+\gamma+1)
\frac{\mu''}{\mu}-
\left(\alpha+\gamma+\alpha\gamma+\frac{3}{4}\right)
\left(\frac{\mu'}{\mu}\right)^2\,,
\ee
where we have used $\beta=1-\alpha-\gamma$.
This is a Riccati equation for the function $(\ln\varsigma(x))'$
given in terms of function $\mu(x)$.

So, for any given PDM $m(x)=1/f^2(x)$, there  exists
function $\varsigma(x)$ such that 
the quantum kinetic term (\ref{fag})
can be presented in the form (\ref{Hetaf}).

\section{Appendix B}\label{AppendixB}

Jacobi elliptic functions are extended 
for the values of the modular parameter $k$ outside the interval 
$(0,1)$ \cite{WW,NIST}.
The $\sn$, $\cn$ and $\dn$  functions are even under 
the change $k\rightarrow -k$.
We also have
\be\label{B1}
\sn\,(z,1/k)=k\,\sn\,(z/k,k)\,,\qquad
\cn\,(z,1/k)=\dn\,(z/k,k)\,,\qquad
\dn\,(z,1/k)=\cn\,(z/k,k)\,,
\ee
and
\be\label{B2}
\sn\,(z,i\hat{k})=k'_1\,\sd\,(z/k'_1,k_1)\,,\qquad
\cn\,(z,i\hat{k})=\cd\,(z/k'_1,k_1)\,,\qquad
\dn\,(z,i\hat{k})=\nd\,(z/k'_1,k_1)\,,
\ee
where 
\be\label{B3}
k_1\equiv \frac{\hat{k}}{\sqrt{1+\hat{k}^2}}\,,\qquad
k'_1\equiv \frac{1}{\sqrt{1+\hat{k}^2}}\,,\qquad
k_1^2+{k'_1}^2=1\,.
\ee
So, for $k=i=i\cdot 1$, we have $\hat{k}=1$, 
$k_1=k'_1=\frac{1}{\sqrt{2}}$, and 
\be\label{B4}
\sn\,(z,i)=\frac{1}{\sqrt{2}}\,\sd\,\left(\sqrt{2}\,z,{1}/{\sqrt{2}}\right)\,,\,\,\,
\cn\,(z,i)=\cd\,\left(\sqrt{2}\,z,{1}/{\sqrt{2}}\right)\,,\,\,\,
\dn\,(z,i)=\nd\,\left(\sqrt{2}\,z,{1}/{\sqrt{2}}\right).
\ee
Note that this is a  special case 
for elliptic functions, for which  $\text{\bf K}(1/\sqrt{2})=\text{\bf K}'(1/\sqrt{2})$,
and the lattice of semi-periods of elliptic functions has additional (rotational in $\pi/2$)
symmetry. It is for this case the elliptic models  we consider can be 
reinterpreted at $k=i$ as those corresponding to a motion of a particle
on Bernoulli lemniscate.

%%%%%%%%%%%%%%%%%%
%%%%%%%%%%%%%%%%%%%%%%%%%%%%%%%%%%%%%

 %%%%%%%%%%%%%%%%%%%%%%%%%%%%%%%%%%%%%%%%%%%%%%%
 %%%%%%%%%%%%%%%%%%%%%%%%%%%%%%%%%%%%%%%%%%%%%%%
  
 %%%%%%%%%%%%%%%%%%%%%%%%%%%%%%%%%%
 %%%%%%%%%%%%%%%%%%%%%%%%%%%%%%%%%%%%%
 %%%%%%%%%%%%%%%%%%%%%%%%%%%%%%%%%%%%%%%%%%%

%%%%%%%%%%%%%%%%%%%%%%%%%%%%%%%%%
%%%%%%%%%%%%%%%%%%%%%%%%%%%%%%%%%%%%%%
%%%%%%%%%%%%%%%%%%%%%%%%%%%%%%%%%%%%%%%%%%%%

\end{document}